\documentclass{article}
\usepackage{emulateapj,epsfig}

\slugcomment{Accepted for publication in ApJ, April 5}

\makeatletter
\newenvironment{inlinetable}{%
\def\@captype{table}%
\noindent\begin{minipage}{0.999\linewidth}\begin{center}\footnotesize}
{\end{center}\end{minipage}\smallskip}

\newenvironment{inlinefigure}{%
\def\@captype{figure}%
\noindent\begin{minipage}{0.999\linewidth}\begin{center}}
{\end{center}\end{minipage}\smallskip}
\makeatother

\begin{document}

\title{Simulations of early structure formation: primordial 
gas clouds}

\author{Naoki Yoshida}
\affil{Harvard-Smithsonian Center for Astrophysics, 60 Garden Street,
Cambridge MA 02138}
\author{Tom Abel}
\affil{Department of Astronomy and Astrophysics, Pennsylvania State University, College
Park, PA 16802}
\author{Lars Hernquist}
\affil{Harvard-Smithsonian Center for Astrophysics, 60 Garden Street,
Cambridge, MA 02138}
\author{Naoshi Sugiyama}
\affil{National Astronomical Observatory Japan, Osawa 2-21-1, Mitaka,
Tokyo 181-8588}

\begin{abstract}

We use cosmological simulations to study the origin of primordial
star-forming clouds in a $\Lambda$CDM universe, by following the
formation of dark matter halos and the cooling of gas within them.  To
model the physics of chemically pristine gas, we employ a
non-equilibrium treatment of the chemistry of 9 species (e$^{-}$, H,
H$^+$, He, He$^{+}$, He$^{++}$, H$_{2}$, H$_{2}^{+}$, H$^{-}$) and
include cooling by molecular hydrogen.  By considering cosmological
volumes, we are able to study the statistical properties of primordial
halos and the high resolution of our simulations enables us to examine
these objects in detail.

In particular, we explore the hierarchical growth of bound structures
forming at redshifts $z\approx 25 - 30$ with total masses in the range
$\approx 10^5 - 10^6 M_{\odot}$.  We find that when the amount of
molecular hydrogen in these objects reaches a critical level, cooling
by rotational line emission is efficient, and dense clumps of cold gas
form.  We identify these ``gas clouds'' as sites for primordial star
formation.  In our simulations, the threshold for gas cloud formation
by molecular cooling corresponds to a critical halo mass of $\approx
5\times 10^5 h^{-1}M_{\odot}$, in agreement with earlier estimates,
but with a weak dependence on redshift in the range $z > 16$.  
The complex interplay between the gravitational formation of dark halos and the
thermodynamic and chemical evolution of the gas clouds compromises
analytic estimates of the critical H$_{2}$ fraction.  Dynamical
heating from mass accretion and mergers opposes relatively inefficient
cooling by molecular hydrogen, delaying the production of star-forming
clouds in rapidly growing halos.

We also investigate the impact of photo-dissociating ultra-violet (UV)
radiation on the formation of primordial gas clouds.  We consider two
extreme cases by first including a uniform radiation field in the
optically thin limit and secondly by accounting for the maximum effect
of gas self-shielding in virialized regions.  For radiation with
Lyman-Werner band flux $J > 10^{-23}$ erg s$^{-1}$ cm$^{-2}$ Hz$^{-1}$
str$^{-1}$, hydrogen molecules are rapidly dissociated, rendering gas
cooling inefficient.  In both the cases we consider, the overall impact
can be described by computing an equilibrium H$_2$ abundance for the
radiation flux and defining an effective shielding factor.

Based on our numerical results, we develop a semi-analytic model of
the formation of the first stars, and demonstrate how it can be
coupled with large $N$-body simulations to predict the star formation
rate in the early universe.

\end{abstract}


\section{Introduction}

The first stars in the Universe almost certainly originated under
conditions rather different from those characterizing present-day star
formation.  Because elements heavier than lithium are thought to be
produced exclusively through stellar nucleosynthesis, the primordial
gas must have been chemically pristine, presumably resulting in stars
of unusually low metallicity.  The recent discovery of an ultra
metal-poor star by Christlieb et al. (2002) suggests that stellar
relics from this era exist even today in our own Galaxy.  Such old
stars offer invaluable information about the history of structure
formation and the chemical composition of the gas in the very early
Universe.

The study of the cooling of primordial gas and the origin of the first
baryonic objects has a long history (e.g., Matsuda, Sato \& Takeda
1969; Kashlinsky \& Rees 1983; Couchman \& Rees 1986; Fukugita \&
Kawasaki 1991; Tegmark et al. 1997).  Within the framework of the
currently favored paradigm for the evolution of structure, i.e.
hierarchical growth by gravitational instability, low mass halos
($\sim 10^6 M_{\odot}$) dominated by Cold Dark Matter (CDM) seed the
collapse of primordial gas within them by molecular hydrogen cooling.
Numerical studies of the formation of primordial gas clouds
and the first stars indicate that this process likely began as early
as $z\approx 30$ (Abel, Bryan \& Norman 2002; Bromm, Coppi \& Larson
2002).  In these simulations, dense, cold clouds of self-gravitating
molecular gas develop in the inner regions of small halos and contract
into proto-stellar objects with masses in the range $\approx 100 - 1000
M_{\odot}$.

While these investigations support the notion that the first stars
were unusually massive, the simulations to date have either mostly
been limited to special cases or ignored the cosmological context of
halo formation and collapse.  In particular, the question of how the
{\it population} of the first luminous objects emerged within a large
cosmological volume has not been explored.

Planned observational programs will exploit future instruments such as
{\it JWST} and {\it ALMA} to probe the physical processes which shaped
the high-redshift Universe.  Among the relevant scientific issues are
the star formation rate at high redshift (e.g., Barkana \& Loeb 2001;
Springel \& Hernquist 2003a), the epoch of
reionization (e.g., Gnedin 2000; Fan et al. 2001; Cen 2002; Venkatesan
et al. 2003; Sokasian et al. 2003; Wyithe \& Loeb 2003a), 
and the fate of high-redshift
systems (White \& Springel 1999).  The statistical properties of early
baryonic objects are of direct relevance to understanding the
significance of the first stars to these phenomena.  In this context,
the key theoretical questions can be summarized as {\it when and where
did a large population of the first stars form?} and {\it how and when
did the Universe make the transition from primordial to ``ordinary''
star formation?}
 
Semi-analytic modeling has often been used to address these questions
qualitatively (see, e.g., Loeb \& Barkana 2001).  Using a spherical
collapse model, Tegmark et al. (1997) estimated the critical H$_{2}$
mass fraction needed for cooling of primordial gas and a corresponding
halo mass scale within which this gas can collapse.  Abel et
al. (1998) and Fuller \& Couchman (2000) later used three-dimensional
simulations to give refined estimates for the minimum collapse mass
(but only for a single or a few density peaks). These results form the
basis for ``minimum collapse mass'' models in which it is assumed that
stars are formed only in halos with mass above a certain threshold.
Using such a treatment, Barkana \& Loeb (2000) and Mackey et
al. (2003) estimated the star formation rate and supernova rate at
high redshift, and Santos, Bromm \& Kamionkowski (2002) computed the
contribution to the cosmic infrared background from massive stars in
the early Universe.  Predictions from these theoretical models are,
however, quite uncertain, because of the relatively crude assumptions
that are used to relate the attributes of luminous objects to those of
dark matter halos.

A few attempts have been made to numerically model early structure
formation in cosmological volumes.  Jang-Condell \& Hernquist (2001)
simulated a cosmological box $1$ Mpc across and found that low mass
($M\sim 10^6M_{\odot}$) dark matter halos at $z\sim 10$ are quite
similar in their properties to larger ones at lower redshifts.
However, they did not include the gas component and hence could not
directly address the nature of the first baryonic objects.
Ciardi et al. (2000) used outputs from $N$-body simulations
to locate star-forming systems in a cosmological volume.
More recently, Ricotti et al. (2002a,b) performed cosmological
simulations including radiative transfer to compute the star formation
history at high redshift.

Feedback processes from the first stars likely played a crucial role
in the evolution of the intergalactic medium and (proto-)galaxy
formation, but the detailed consequences of these effects remain
somewhat uncertain.  Radiation can produce either negative
feedback, by dissociating molecular hydrogen via Lyman-Werner
resonances (Dekel \& Rees 1987; Haiman, Abel \& Rees 2000; Omukai \&
Nishi 1998, 1999), or positive feedback from X-rays which can promote
H$_2$ production by boosting the free electron fraction in distant
regions (Haiman, Rees \& Loeb 1996; Oh 2001).  It is not clear whether
negative or positive feedback dominates.  Machacek, Bryan \& Abel
(2001) examined the the former using numerical simulations which
included an H$_{2}$ photo-dissociating radiation field of constant
flux in the optically thin limit.  Cen (2002) emphasized the positive
impact of an early X-ray background on the formation of the first
stars and discussed the possibility that the Universe could have been
re-ionized at an early epoch by Population III objects alone 
(see also Wyithe \& Loeb 2003a, 2003b).  
Using three-dimensional adaptive mesh refinement (AMR) simulations, Machacek
et al. (2003) further argued that the net effect of an X-ray background
on gas cooling is milder than one naively expects from simple analytic
estimates.

The formation of the first stars, the evolution of early cosmic
radiation fields, and the thermal properties of the high-redshift
intergalactic medium (IGM) are closely linked, and it is likely that
semi-analytical studies of the formation of pre-galactic objects are
limited by the complex {\it cross-talk} between them.  Clearly,
high-resolution simulations in a proper cosmological context are
needed to advance our understanding of the details of first structure
formation in the early Universe.

In the present paper, the first in a series on early structure
formation, we study the cooling and collapse of primordial gas in dark
halos using high resolution cosmological simulations.  We evolve the
nonequilibrium rate equations for 9 species and include the relevant
gas heating and cooling in a self-consistent manner.  From a
large sample of dark halos, we determine conditions under which the
first baryonic objects form.  We show that ``minimum collapse mass''
models are a poor characterization of primordial gas cooling and gas
cloud formation, because these processes are significantly affected 
by the
dynamics of gravitational collapse.  We also examine the influence of
H$_{2}$ dissociating radiation in the form of a uniform background
field in the optically thin limit as well as by approximately
accounting for gas self-shielding.  We quantify the overall negative
effect of photo-dissociating radiation using both the numerical
results and analytic estimates for the efficiency of gas cooling.
Based on the simulation results, we develop a semi-analytic model to
describe the formation of star-forming gas clouds in dark matter
halos.  To this end we adopt a simple star-formation law and compute
the global star-formation rate using a large $N$-body simulation.

The paper is organized as follows. In section 2 we describe our
numerical simulations.  Section 3 presents general results of the
simulations, ignoring radiation.  Sections 4, 5 and 6 give basic
properties of the dark matter halos found in our simulations.  We
discuss the impact of far UV radiation on primordial gas cooling in
section 7.  We describe a semi-analytic model for early star formation
and its applications in section 8.  Concluding remarks are given in
section 9.

\newpage

\begin{inlinetable}
\begin{center}
\caption{Simulation parameters}
\begin{tabular}{ccccc}
\tableline
\tableline
Run & $N_{\rm tot}$ & L ($h^{-1}$kpc) & $m_{\rm gas}$ ($h^{-1} M_{\odot}$)
& $l_{\rm s}$ ($h^{-1}$pc) \\
\tableline
A & 2$\times 288^{3}$ & 600 & 100.0 & 54 \\
\tableline
B & 2$\times 216^{3}$ & 300 & 29.6  & 36 \\
\tableline
C1 & 2$\times 144^{3}$ & 300 & 100.0 & 54\\
\tableline
C2 & 2$\times 144^{3}$ & 300 & 100.0 & 54\\
\tableline
DM & $324^{3}$ & 1600 & ($m_{\rm DM}$) 10000.0  & 200\\
\tableline
\end{tabular}
\end{center}
\end{inlinetable}

\section{The $N$-body/SPH simulations}

We use the parallel $N$-body/SPH solver GADGET (Springel, Yoshida \&
White 2001), in its ``conservative entropy'' formulation (Springel \&
Hernquist 2002).  We follow the non-equilibrium evolution of nine
chemical species (e$^{-}$, H, H$^+$, He, He$^{+}$, He$^{++}$, H$_{2}$,
H$_{2}^{+}$, H$^{-}$) using the method of Abel et al. (1997) and we
employ the cooling rate of Galli \& Palla (1998) for molecular
hydrogen cooling.  The time stepping method employed in the SPH
simulations is described in the Appendix. 

The largest of our chemo-hydrodynamic simulations (Run A) employs 48 million
particles in a periodic cosmological box of $600h^{-1}$kpc on a side.
We consider a conventional $\Lambda$CDM cosmological model with matter
density $\Omega_{0}=0.3$, cosmological constant $\Omega_{\Lambda}=0.7$
and present expansion rate $H_{0}=70$km s$^{-1}$Mpc$^{-1}$. The baryon
density is $\Omega_{\rm b}=0.04$.  The initial power spectra for the
baryonic and dark matter components are accurately computed from the
Boltzmann code of Sugiyama (1995), in which the pressure term of 
baryon perturbations is taken into account. 
The initial power spectrum is normalized by setting $\sigma_{8}=0.9$,
and all of our simulations are started at $z=100$.  
%
The initial ionization fraction was computed using
RECFAST (Seager, Sasselov \& Scott 2000) and was set to be $x_{e} =
2.984\times 10^{-4}$ for the $\Lambda$CDM universe we adopt.  
Details of the set-up of the initial conditions will be presented
elsewhere (Yoshida, Sugiyama \& Hernquist 2003).

The basic simulation parameters are listed in Table 1. There, $L$
is the simulation box side length, $m_{\rm gas}$ is the mass per gas
particle, and $l_{\rm s}$ is the gravitational softening length.  
Run B was carried out with a higher mass and spatial resolution to check the
convergence of our numerical results.  It started from the same
initial matter distribution as that of C1 on large scales.  
Runs C1 and C2 differ only in the assigned
phase information and fluctuation amplitudes in the initial random
Gaussian fields. They are used to test how sample variance of the
initial density field affects the final results. 
We carried out a simulation with dark matter only, denoted ``Run DM'', to
construct halo merger histories to construct the semi-analytic model
described in section 8.   We continued the simulations until about 
100 million years after the first bound object formed.  
Radiation from the first object(s)
should, in principle, be included because it affects the chemical and
thermodynamic evolution of the surrounding IGM in a large fraction of
our simulated regions.  Nevertheless,
we do not take such radiative effects into account in the first series
of our simulations, in order to isolate other dynamical effects on
primordial gas cloud formation.  We carry out the same set of
simulations with UV radiation in the Lyman-Werner bands
and examine the global effect of photo-dissociation in section 7.

During the simulations, we save 64 snapshots of the particle data
spaced logarithmically in cosmic expansion parameter from redshift
$z=100$ to $z=14$. We use these outputs to identify cold, dense gas
clouds, and to construct dark matter halo merger trees.  We locate
dark matter halos by running a friends-of-friends (FOF) groupfinder
with linking parameter $b=0.164$ (Jenkins et al. 2001) in units of mean particle separation,
and discard the groups which have less than 100 dark matter particles.
We define the virial radius $R_{\rm vir}$ of a halo as the radius of
the sphere centered on the most bound particle of the FOF group having
overdensity 180 with respect to the critical density.  The virial mass
$M_{\rm vir}$ is then the enclosed mass (gas + dark matter) within
$R_{\rm vir}$.

\vspace*{3mm}
\begin{inlinefigure}
\resizebox{8cm}{!}{\includegraphics{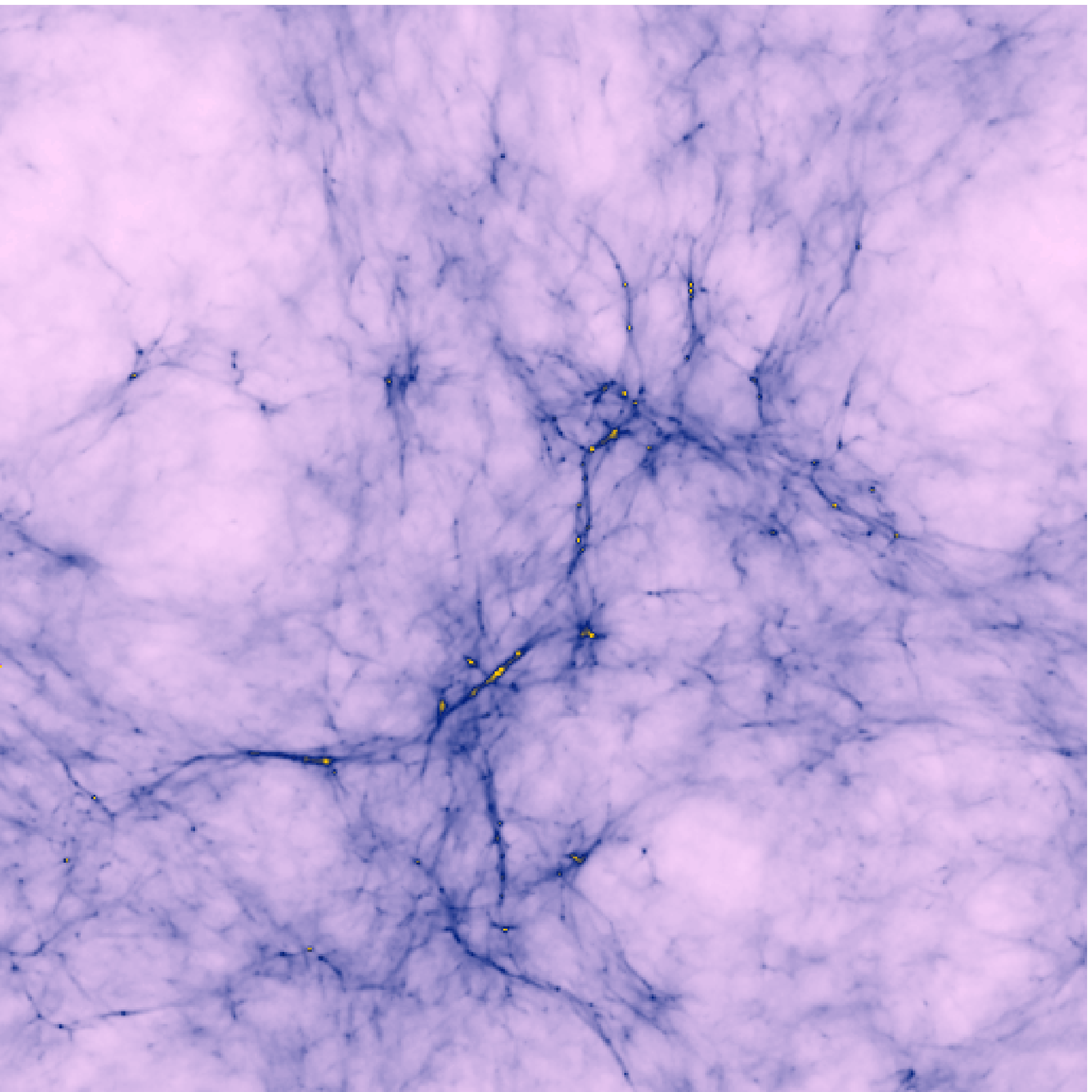}}
\caption{The projected gas distribution in the simulation box for Run A
at $z=17$. The cooled dense gas clouds appear as bright spots at the 
intersections of the filamentary structures. \label{plot1}}
\end{inlinefigure}

\section{The minimum collapse mass}

For the halos identified in the simulations, we measure the mass of
gas which is cold ($T < 0.5 T_{\rm vir}$) and dense ($ n_{\rm H} >
5\times 10^2 $cm$^{-3}$).  Once the gas starts to cool, clouds of
molecular gas grow rapidly at the centers of halos and their masses
exceed the characteristic Jeans mass $M_{\rm J}\sim3000 M_{\odot}$ for
the typical temperature $T\sim 200$ K and density $n_{\rm H} \sim 10^3
$cm$^{-3}$ of the condensed primordial gas.  Hence, they are expected
to be sites for active star formation.  Hereafter, we refer to such
cold, dense gas clumps as ``gas clouds.''  
Since a halo can host more than one gas clouds, we run a FOF groufinder 
independently to the gas particles with a very small linking parameter
$b=0.05$. In this manner we can separate groups of dense gas particles.
We then discard groups of gas particles that do not satisfy the
above criteria of cold, dense gas.
By checking the locations of the selected groups in all the outputs, 
however, we found no halos which host more than one gas clouds
in this particular simulation.
Figure \ref{plot1} shows the
projected gas distribution in the simulation box of Run A at
$z=17$. The bright spots are the primordial gas clouds.  There
are 31 gas clouds in the simulated volume.  It is important to note
that, whereas most of them are strongly clustered in 
\begin{inlinefigure}
\resizebox{8.5cm}{!}{\includegraphics{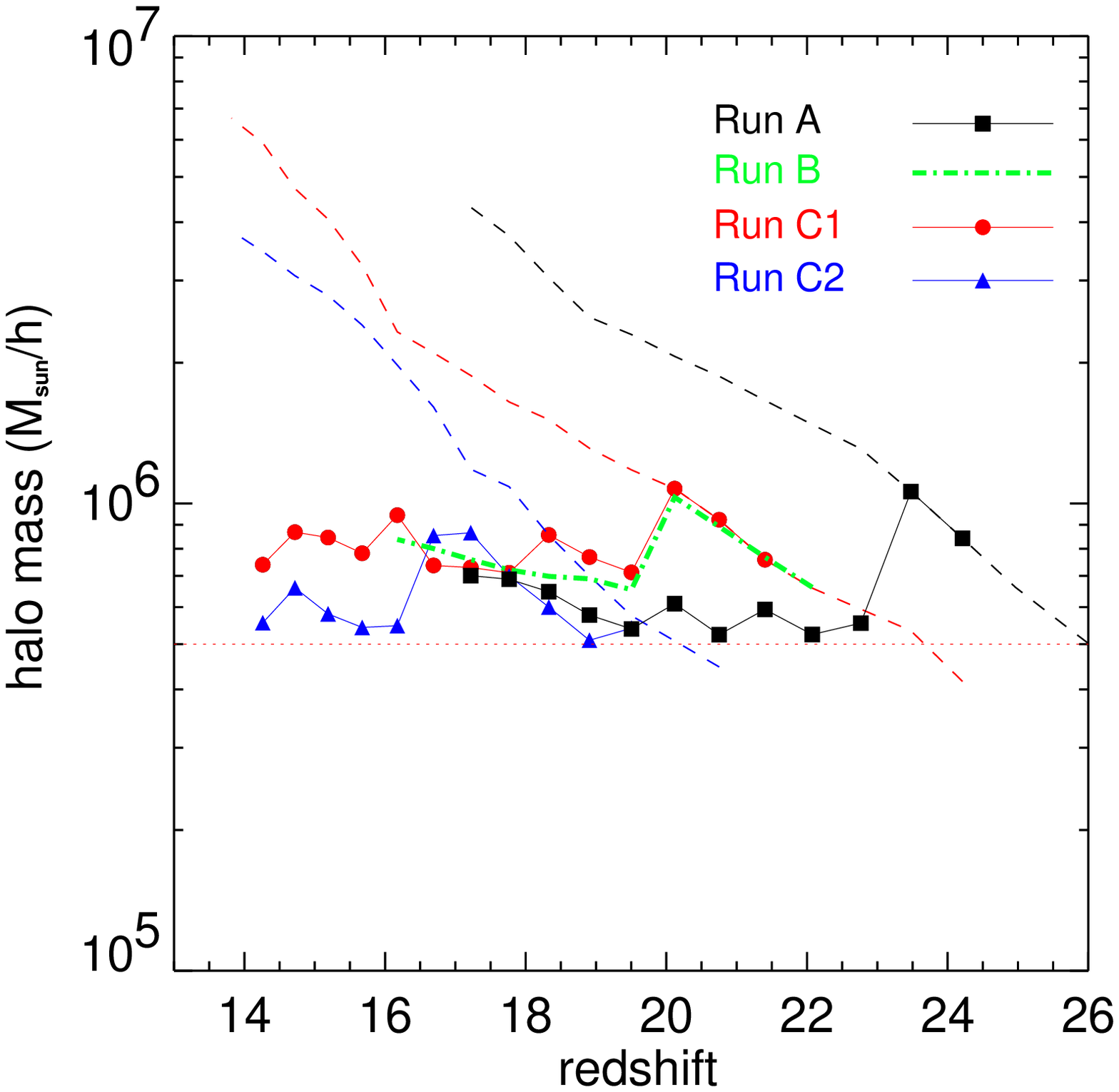}}
\caption{The minimum mass of the halos that host cold gas clumps.
The solid lines with symbols indicate the minimum mass at the output 
redshifts, and the dashed lines show the 
mass evolution of the most massive halo in each run. 
\label{plot2}}
\end{inlinefigure}
\\
high density
regions, reflecting the clustering of the underlying dark matter,
some gas clouds are found in less dense, isolated regions.

In Figure \ref{plot2} we plot the minimum mass of the halos that host
gas clouds at each output redshift.  It approximates the evolution of
the minimum mass of the star-forming systems.  In the figure, we also
show the evolution of the most massive halo in each run.  The apparent
earlier formation epoch of the first bound object in Run A is simply
due to finite volume effects.  Run A simulated a volume 8 times 
larger than
the others, and thus it contained a higher-$\sigma$ density
fluctuation than Runs C1 or C2.  Cosmic variance also explains the
difference in the minimum mass between Runs C1 and C2. The assigned
initial power spectrum for Run C1 had, by chance, somewhat larger
amplitudes than for Run C2 on the largest scales.  We checked that the
dark halo mass functions are noticeably different at large mass scales
between the two runs.  On the other hand, excellent agreement is found
in the minimum mass scale between the high resolution Run B
(dot-dashed line) and low resolution Run C1 (filled circles). Our
result appears to be converged on mass scales which our simulations
probe.

Figure \ref{plot2} clearly shows that the minimum collapse mass scale
lies at $M_{\rm cr} = 5\times 10^5 h^{-1}M_{\odot}$, with only a weak
dependence on redshift in the range plotted.  Our result agrees
reasonably well with that of Fuller \& Couchman (2000), who
carried out three-dimensional simulations for single high-$\sigma$
density peaks.  The weaker redshift dependence found by us reflects
the fact that we define the minimum collapse mass using a large sample
of halos formed in various places in the simulation volume, rather
than for a single or a few objects in high density regions.  Our result
is also roughly consistent with that of Machacek et al. (2001; 2003),
who obtain a smaller value for the minimum mass, $\sim 2-3 \times 10^5
h^{-1}M_{\odot}$.

An important quantity that determines the onset of gas cooling is the
fraction of hydrogen molecules, $f_{\rm H_{2}}$.  In Figure \ref{plot3}
we plot $f_{\rm H_{2}}$ against the virial temperature for halos 

\begin{inlinefigure}
\resizebox{8.5cm}{!}{\includegraphics{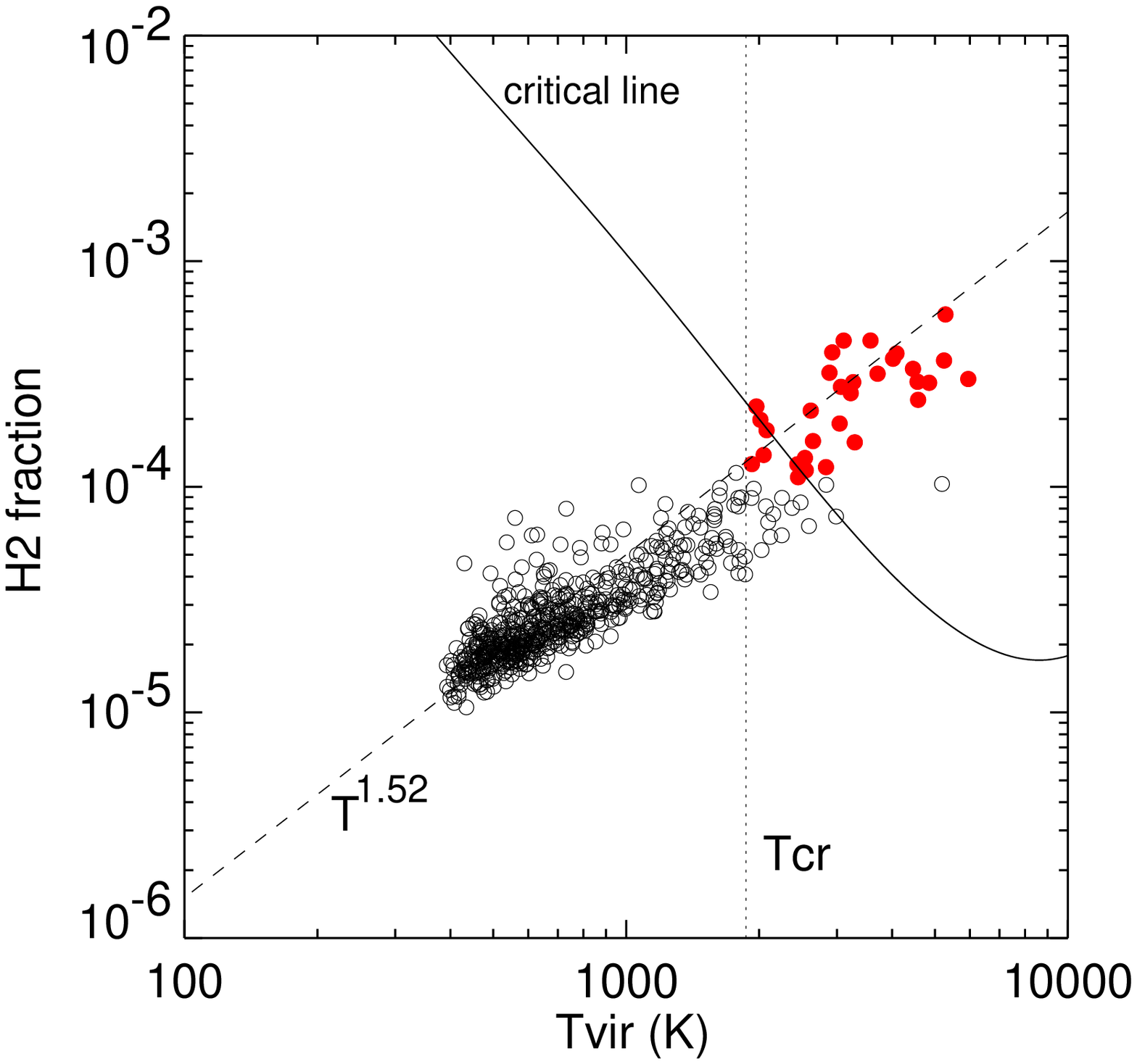}}
\caption{The mass weighted H$_{2}$ fraction versus virial temperature 
for the halos that host
gas clouds (filled circles) and for those that do not (open circles)
in Run A at $z=17$. The virial temperature is related to the halo mass by equation 
(1).
The solid curve is the H$_{2}$ fraction needed to cool the gas
at a given temperature and the dashed line is the asymptotic H$_{2}$
fraction.
\label{plot3}}
\end{inlinefigure}
\\
in Run A at $z=17$.  We compute the virial temperature for the halo mass
using
\begin{eqnarray}
T &=& 1.98\times 10^4 \left(\frac{\mu}{0.6}\right) 
	\left(\frac{M_{\rm halo}}{10^8 h^{-1}M_{\odot}}\right)^{2/3} \\ \nonumber
  &\;& \left(\frac{\Omega}{\Omega_{z}}\frac{\Delta}{18\pi^2}\right)^{1/3}
	\left(\frac{1+z}{10}\right) \mbox{K},
\label{eq_Tvir}
\end{eqnarray}
where $\mu$ is the mean molecular weight and $\Delta$ is the collapse
overdensity.  Filled circles represent the halos harboring a cold
dense gas cloud, while open circles are for the others.  The
solid line is an analytical estimate of the H$_{2}$ fraction needed to
cool the gas, which we compute {\it a l\`{a}} Tegmark et
al. (1997). 
Briefly, we compute the characteristic cooling time
of a gas with density $\rho$ and temperature $T$ as 
$t_{\rm cool} = k_{\rm B}T/\rho \Lambda (\rho, T)$
where $\Lambda (\rho, T)$ is the cooling rate due to molecular hydrogen
rotational line transitions, 
and determine the critical molecular hydrogen abundance with
which the gas can cool within a Hubble time.
Note that we use the cooling function of Galli \& Palla
(1998) for our simulations and for this estimate. The dashed line is
the asymptotic molecular fraction of a gas in a transition regime when
electron depletion makes the production of hydrogen molecules
ineffective.  Then the molecular fraction scales as $f_{\rm H_2}
\propto T^{1.52}$ (see eq. [17] of Tegmark et al. 1997)

In Figure \ref{plot3}, halos appear to be clearly separated into two
populations; those in which the gas has cooled (top-right), and the
others (bottom-left).  Our analytic estimate indeed agrees very well
with the distribution of gas in the $f_{\rm H_{2}}$ - $T$ plane.
We emphasize, however, that this apparent agreement should not be
interpreted as the model precisely describing the gas evolution.
Also, the analytic model itself is expected to be accurate only to
within some numerical factor.

Although the H$_{2}$ fraction primarily determines whether the gas in
halos can cool or not, there are some halos within which gas clouds
have not formed despite the high gas temperatures (open circles
with $T > T_{\rm cr}$).  At $z=17$, about 30\% of the massive halos
are ``deficient'' in this manner.  Similar features are also found in
the result of the AMR simulation of Machacek et al. (2001, their
Fig. 3).  In the next section we further examine what prevents the gas
from cooling and collapsing in these halos.

\section{Halo formation history}

Halos in CDM models grow hierarchically through merging and the
accretion of smaller objects.  The complex and violent formation
processes of dark halos affect the thermal and chemical evolution of
the gas within them.  To addess this, we study the dynamical influence
of dark matter on gas cloud evolution using halo merger histories.
We identified a total of 635 dark halos in Run A at $z=17$.  For all
the halos, we traced their progenitors in earlier outputs to construct
merger trees.  In Figure \ref{plot4} we plot the mass evolution of a
subset of halos that host gas clouds (top-left panel) and of another
subset of halos that do not host gas clouds (top-right panel).  In the
top-left panel, we mark trajectories by filled circles when they host
gas clouds. The figures show a clear difference between the two
subgroups in their mass evolution. Most of the halos in the top-left
panel experience a gradual mass increase since the time their masses
exceeded $M_{\rm cr}$, whereas those plotted in the top-right panel
have grown rapidly after $z\sim20$. It appears that the gas in halos
that accrete mass rapidly (primarily due to mergers) is unable to cool
efficiently.

Rapid mass accretion and mergers dynamically heat the gas when halos
form, causing it to become hot and rarefied, rather than allowing it
to radiatively cool and condense.  This is the situation first
considered by White \& Rees (1978) in the context of hierarchical
galaxy formation.  The simplest model to describe the evolution of
radiative gas assumes that the gas cools when the characteristic
cooling time is shorter than the dynamical time.  This scenario has been
used to estimate the minimum mass scale for galaxies.  The gas cooling
rate due to atomic hydrogen and helium associated with galaxy
formation has a particular behavior in
it decreases sharply below $T = 10^4$ K by many orders of magnitude, which effectively prevents the
cooling of gas in systems with low virial temperatures.  Hence, the
cooling criterion $t_{\rm cool} < t_{\rm dyn}$ simply sets a definite
minimum mass scale (at a given redshift) for galaxy formation.

The situation we consider here is clearly different, because molecular
hydrogen cooling is a much {\it less} efficient process and has a
weaker dependence on temperature.  More important, for this mechanism
to be effective, a certain number of hydrogen molecules must first be
produced, because the residual H$_{2}$ abundance in the early Universe
is negligible.  Figure \ref{plot4} illustrates these features.  In the
bottom panels, we plot the evolution of the molecular hydrogen
fraction and the mean gas mass weighted temperature for the same halos
as in the top panels.  Most of the trajectories in the bottom-right
panels show a common feature: the temperature rises with little
increase in the molecular hydrogen fraction.

We can understand this as follows.  Consider an equal-mass merger
where two halos each with a mass $5\times 10^5 h^{-1}M_{\odot}$ merge
to form an object of mass $10^6 h^{-1}M_{\odot}$ at $z=20$.  The H$_2$
fraction of the gas in the two halos is

\begin{inlinefigure}
\vspace*{0.5cm}\ \\
\resizebox{4.3cm}{!}{\includegraphics{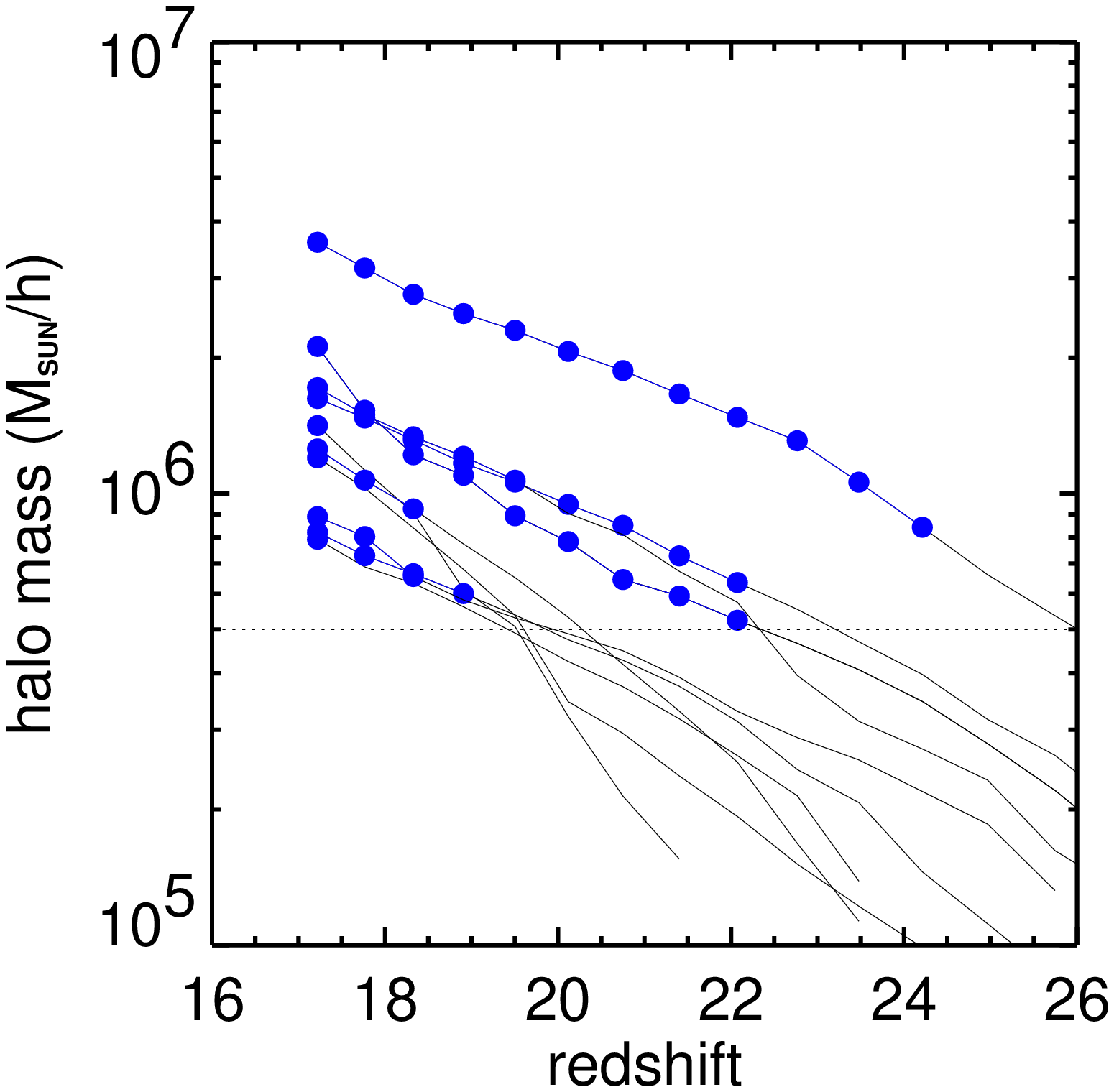}}
\resizebox{4.3cm}{!}{\includegraphics{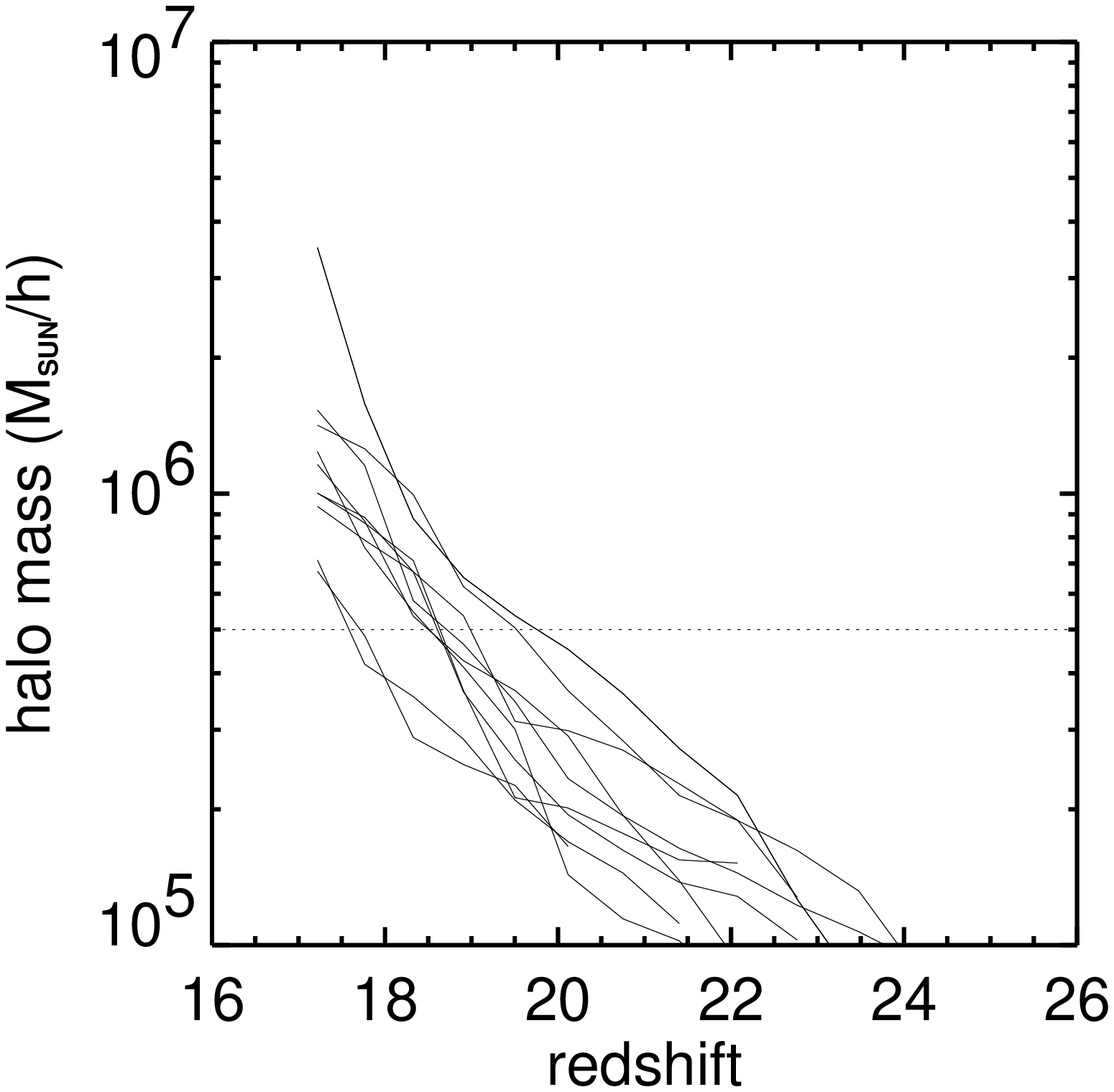}}\\
\vspace{0.2cm}\ \\
\resizebox{4.3cm}{!}{\includegraphics{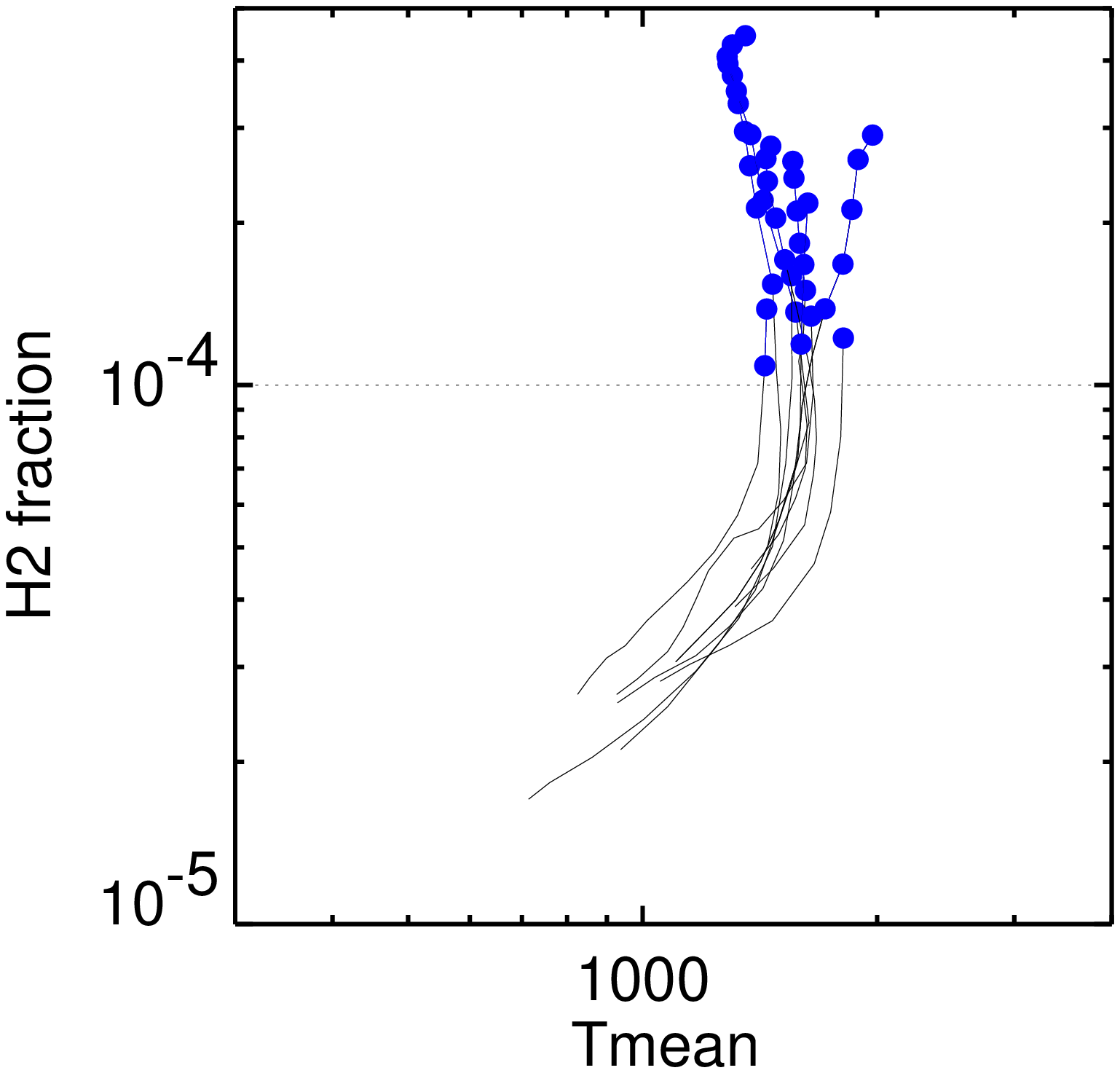}}
\resizebox{4.3cm}{!}{\includegraphics{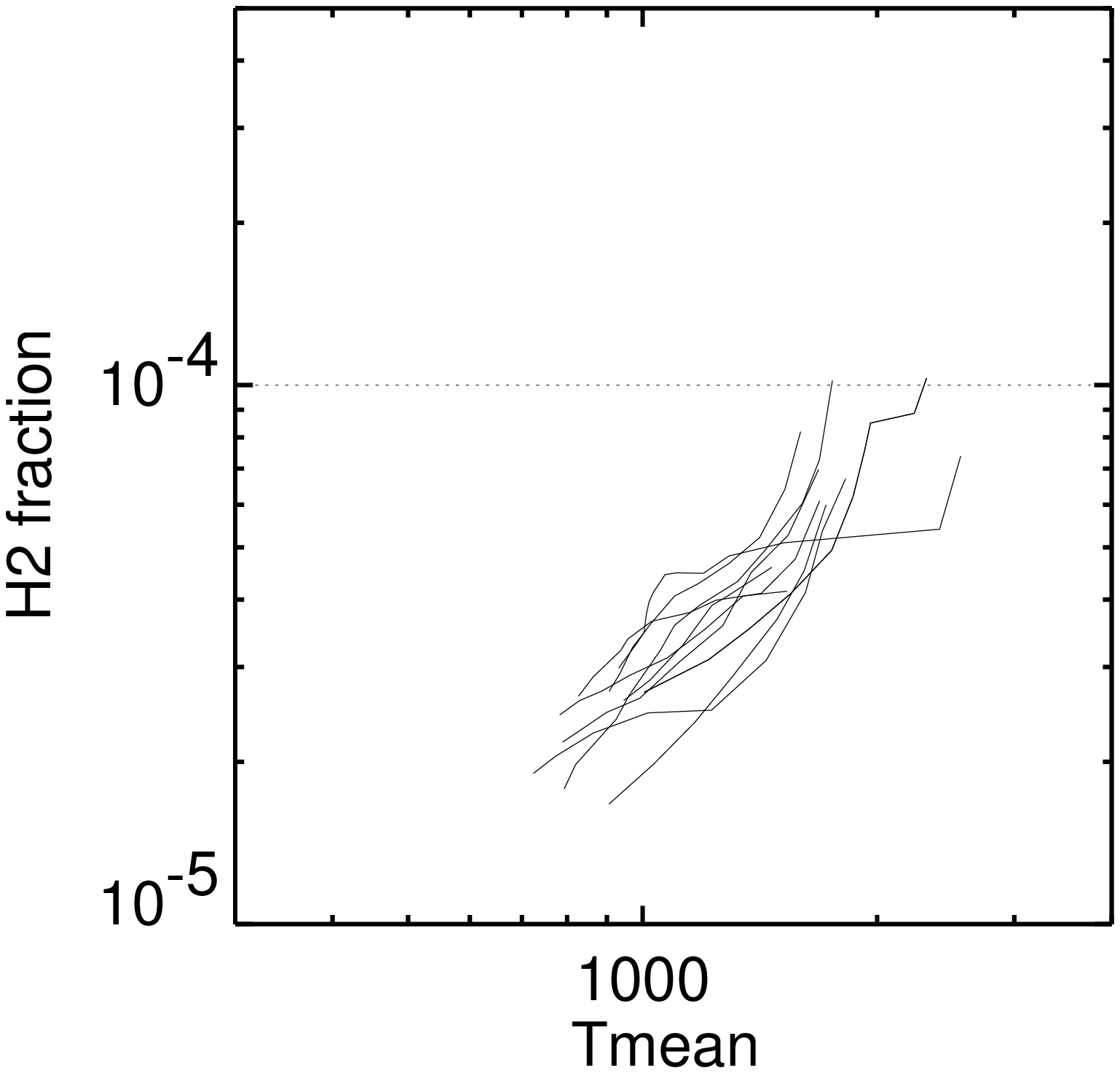}}
\caption{Top panels: The mass evolution of the halos that host gas clouds at z=17 (left)
and those that do not (right) for Run A. 
Bottom panels: The molecular hydrogen fraction is plotted against
the mean gas-mass-weighted temperature for the same halos as in the top panels.
\label{plot4}}
\end{inlinefigure}
\\
predicted to be about $8.0\times 10^{-5}$ (see the dashed line in Figure \ref{plot3}).  After
virialization, the gas temperature becomes $T_{\rm
vir}\sim 2300$K, and our estimate for the H$_2$ fraction needed for
the gas to cool is about $2\times 10^{-4}$, about a factor of 2.5 
larger
than the progenitor gas.  In this case, the production of molecular
hydrogen, the coolant, must precede cooling.  Only when a further
temperature increase or an increase in the molecular fraction 
brings the gas into
the region above the critical line in the $f_{\rm H_2} - T$ plane
can the
gas cool efficiently, unless significant heating occurs
during this cooling phase.  The molecular hydrogen formation time scale
for this typical halo is estimated to be
\begin{equation}
t_{\rm H_2}=\frac{n_{\rm H_2}}{k_{\rm H^{-}} n_{\rm H} n_{\rm e}} \approx 30 {\rm Myrs}, 
\end{equation}
where $k_{\rm H^{-}}$ is the reaction coefficient of H$^{-}$ formation via
H + e$^{-}$ $\rightarrow$  H$^{-}$ + $\gamma$.
The H$_2$ formation timescale is comparable to the dynamical time 
for this halo.

It is interesting that the most massive halo plotted in the top-right
panel of Figure \ref{plot4} has a mass of $3.5\times 10^6 h^{-1}M_{\odot}$ at
$z=17$.  The mass growth of the halo is so rapid below $z = 20$, 
when it had a mass of $5\times 10^5 h^{-1}M_{\odot}$, that 
the gas within it could not cool to form a dense gas cloud.
It has been instead continuously {\it heated} dynamically.
We can quantify this dynamical effect using our simulations.
We trace the progenitors of the massive 
($M>7\times 10^5 h^{-1}M_{\odot}$) halos in Run A identified at $z=17$,
and compute recent mass
accretion rates from $\Delta M/\Delta z = (M(z_{2})-M(z_{1}))/(z_{2}-z_{1})$,
where we choose $z_{1}=18.5$ and $z_{2}=17$. 
In Figure \ref{plot5} we plot the measured mass growth rates against
the halo masses.
Filled circles represent the
halos that host gas clouds at $z=17$, and open circles are for
those that do not. 
We derive a critical mass growth rate by equating the heating rate
to the 
\begin{inlinefigure}
\resizebox{8cm}{!}{\includegraphics{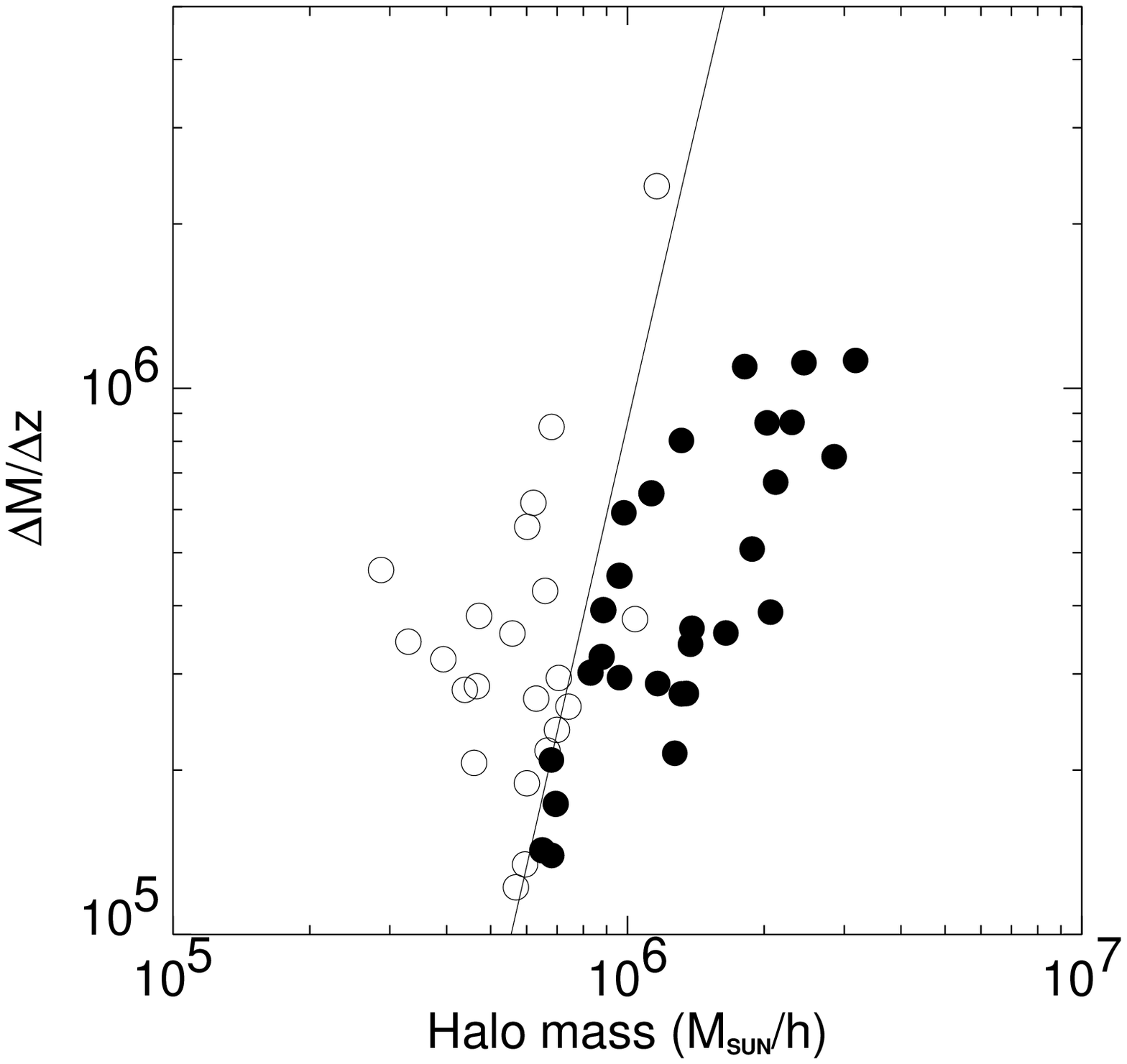}}
\caption{The mass growth rate versus halo mass
for halos at z=18.5 in Run A.
The mass growth rates are computed from  $\Delta M/\Delta z = (M(z_{2})-M(z_{1}))/(z_{2}-z_{1})$,
where $z_{1}=18.5$ and $z_{2}=17$. Note that, according to our definition, the mass increase
per unit redshift can be larger than the halo mass itself at $z_1$, if the halo's
descendant is more massive (up to by a factor of two) due to successive mergers during the redshift interval
considered. The solid line shows the critical {\it instantaneous} mass growth rate 
computed from the dynamical heating rate that balances the estimated cooling rate. 
\label{plot5}}
\end{inlinefigure}
\\
molecular hydrogen cooling rate from
\begin{eqnarray}
\frac{dQ_{\rm dyn. heat}}{dt}&\equiv&\frac{k_{\rm B}}{\gamma-1}\frac{dT}{dt}\\ \nonumber
&=& \Lambda_{\rm H_{2}}(T) \times f_{\rm H_{2}}\\ \nonumber
& & (\mbox{cooling rate per hydrogen atom}),
\label{eq_heat}
\end{eqnarray}
and relate the increase in virial temperature to the mass growth rate by
\begin{equation}
\frac{dT}{dt}=\alpha M^{-1/3}\frac{dM}{dt},
\label{eq_heat2}
\end{equation}
where the coefficient $\alpha$ is computed from equation (1) at a given time.
Strictly speaking, the temperature of the gas in a halo could be
different from the halo's virial temperature. We have carried out an
additional simulation with non-radiative gas starting from the same
initial conditions as Run B (the highest resolution simulation) and
found that the mean gas-mass weighted temperature is indeed close to
the halo's virial temperature for almost all the halos, with
deviations smaller than a factor of two (see also Figure 2 of Machacek et
al. 2001).  Therefore, we may safely assume that the gas temperature is
close to the virial temperature before cooling occurs. 

In Figure \ref{plot5}, the solid line is our analytic estimate for the critical
mass growth rate.  Above the solid line, the mass growth rate is so
large that dynamical heating acts more efficiently than
cooling by molecular hydrogen. Thus, in such halos, gas cloud
formation is effectively delayed or prevented.  Note the steepness of
the critical mass growth rate, which reflects the slope of the
molecular hydrogen cooling rate $\Lambda (T) \propto T^{1.5}$ in the
temperature and density range of interest. For a halo with mass
$5\times 10^5 h^{-1}M_{\odot}$ at $z\sim 20$, even a 20\% mass
increase per unit redshift results in net heating of the gas.
The critical
mass growth rate 

\begin{inlinefigure}
\resizebox{6.5cm}{!}{\includegraphics{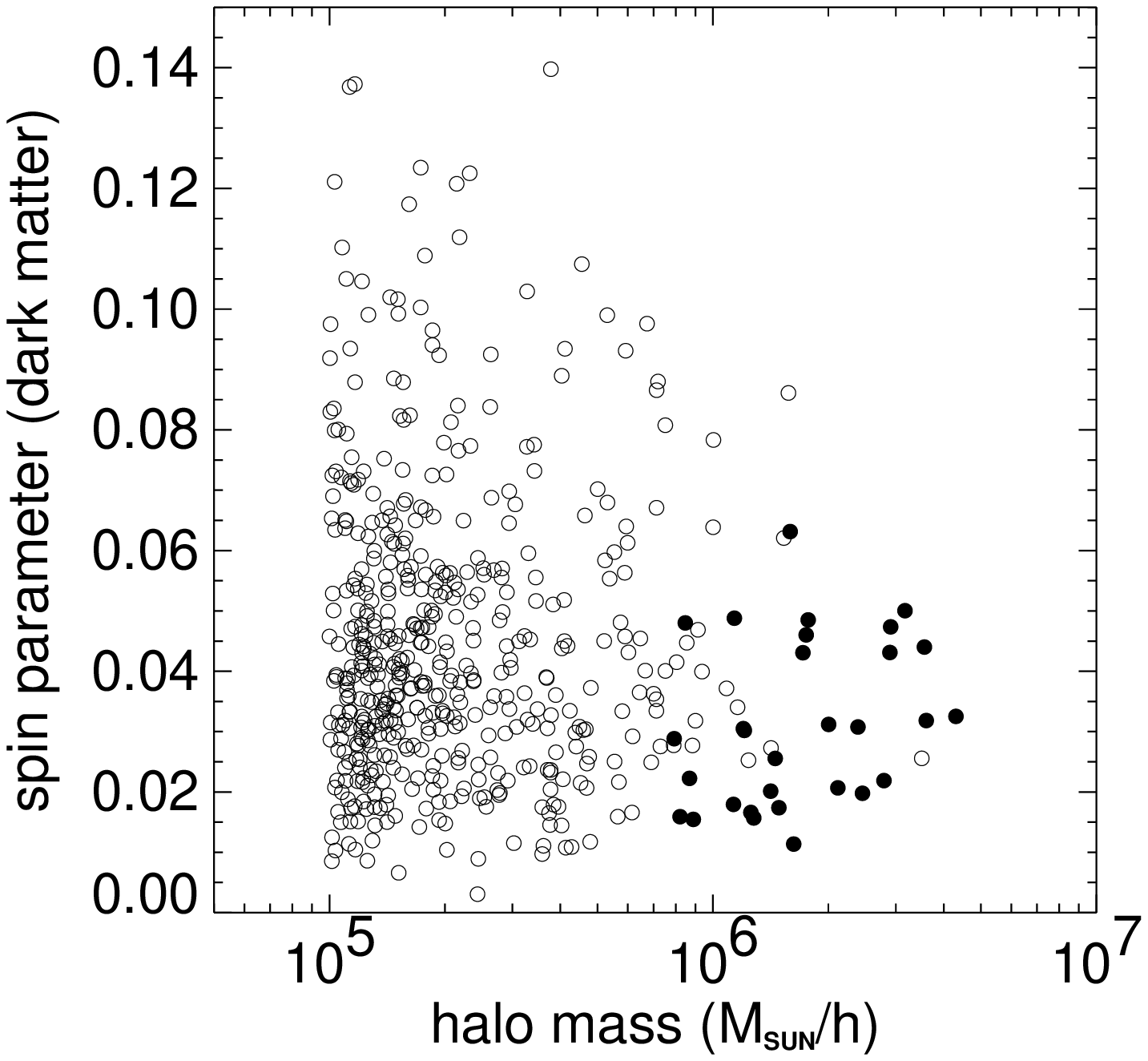}}\\
\resizebox{6.5cm}{!}{\includegraphics{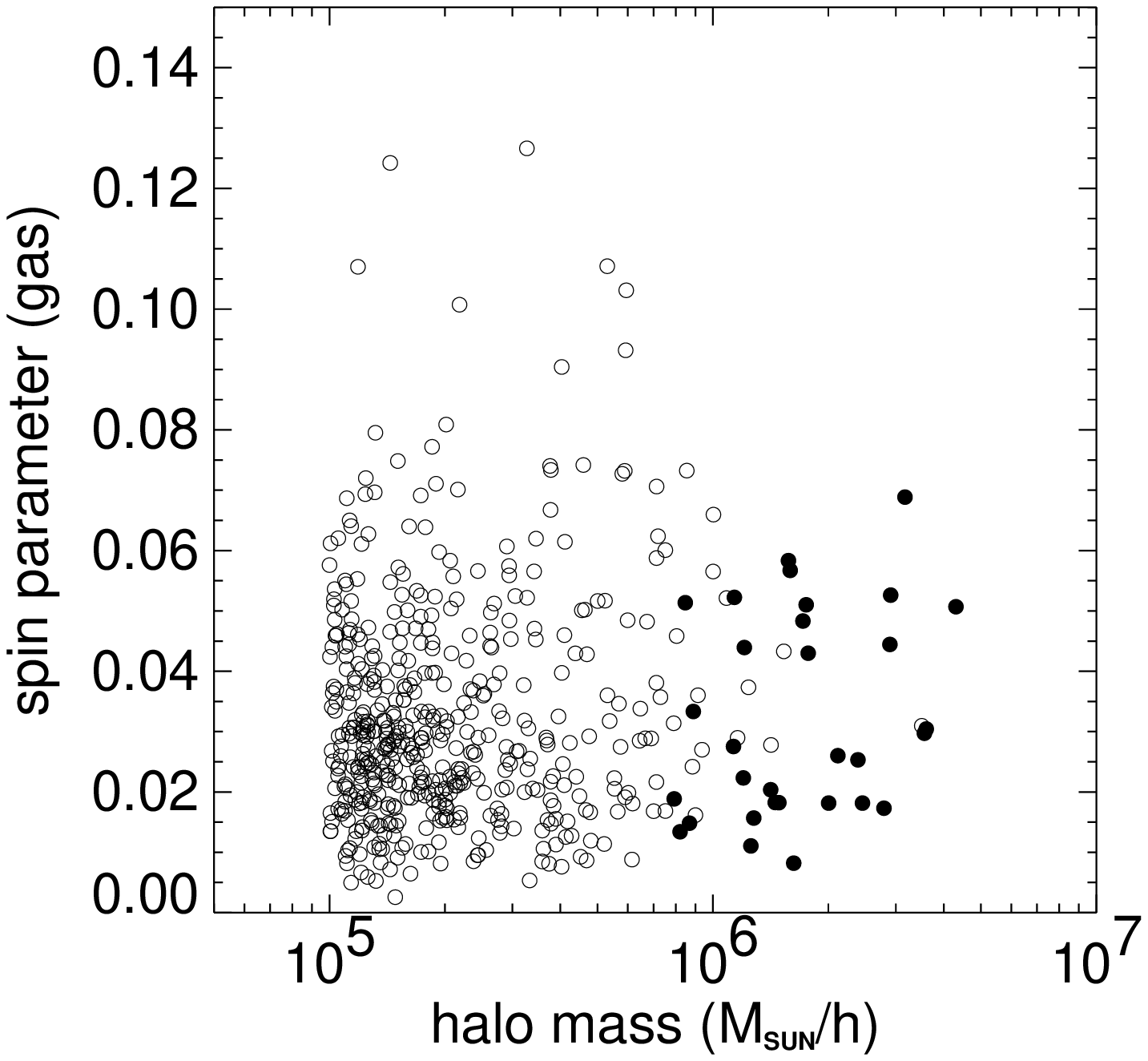}}
\caption{Spin parameters of halos are plotted against
their masses, for the dark matter component (top) and for the gas (bottom)
in our largest run (Run A) at $z=17$.
We plot halos with mass greater than $10^5 h^{-1}M_{\odot}$.  Filled and
open circles indicate halos with and without gas clouds, respectively.
\label{plot6}}
\end{inlinefigure}
\\
therefore sets a natural lower limit for
primordial gas cloud formation at a mass scale $\sim 5\times 10^5
h^{-1}M_{\odot}$, in good agreement with the result shown in Figure \ref{plot2}.

\section{The angular momenta of dark matter and gas}

Recent numerical studies by Bromm et al. (2002) show that the initial
angular momentum of primordial gas may determine the properties of
the first star-forming clouds and possibly of the stars
themselves.  By
pre-assigning the system's initial angular momentum and following its
evolution, Bromm et al. (2002) found that a disk-like structure is
formed in high spin systems and that the gas subsequently
fragments.  It is therefore interesting to ask whether such
high spin systems are indeed produced in cosmological simulations.  We
measure the spin parameters of the dark matter component and of the
gas within them.  We follow the definition of Bullock et al. (2001):
\begin{equation}
\lambda = \frac{j}{\sqrt{2} R_{\rm vir} V_{\rm vir}},
\end{equation}
where $j$ is the specific angular momentum of each component (dark
matter or gas) and 
$V_{\rm vir} = \sqrt{G M_{\rm vir}/R_{\rm vir}}$ is the circular
velocity at the virial radius.

In Figure \ref{plot6}, we plot the spin parameters of the dark matter
and of the gas against the halo mass for Run A at $z=17$. 
For the gas component, we included all the gas (hot + cold) particles
within the virial radius.
The distribution of the dark matter spin parameters is quite similar to
that of both high mass halos (van den Bosch et al. 2001) and small
halos (Jang-Condell \& Hernquist 2001).  The spin parameter
distribution for the dark matter is well fitted by the lognormal
function
\begin{equation}
p(\lambda) d\lambda = \frac{1}{\sqrt{2\pi}\sigma_{\lambda}}
\exp\left[-\frac{\ln^2 (\lambda/\bar{\lambda})}{2\sigma^{2}_{\lambda}}\right]
\frac{d\lambda}{\lambda},
\end{equation}
 with $\bar{\lambda}=0.035$ and $\sigma_{\lambda}=0.54$.  We
also note that the spin vectors of dark matter and the gas are not
closely aligned, with a median deviation angle of $\approx 30$
degrees, in good agreement with the results of van den Bosch et
al. (2001) for higher mass halos.

In Figure \ref{plot6}, the spin parameters appear relatively smaller
for halos with gas clouds
than for the entire halo population, because gas clouds form only
in halos at the high-mass end.  Bromm et al's result suggests that in
systems with a spin parameter as large as 0.06, gas clouds eventually
flatten to form a rotationally supported disc.  We find only two 
star-forming halos
in which either the gas or dark matter spin parameters are greater than
0.06.  Although rare, such objects do form in the CDM model.  It is
important to point out, however, that the spin vectors of the gas and
dark matter are {\it not} usually aligned.  This confirms the
importance of setting up simulations in a proper cosmological context.
Intriguingly, Vitvitska et al. (2002) argue that 
halos which have experienced recent
major mergers tend to have high spin parameters. 
This may explain the overall trend in Figure \ref{plot6} 
that gas clouds are preferentially found in low-spin halos.

It is still a difficult task to measure the spin parameters of the
formed dense gas clouds accurately, because of resolution limitations.
To address the question of the exact shape and the size of the final
gas clump, we require a substantially higher resolution simulation.

\section{Gas cloud evolution}

The cooling and condensation of the gas within a dark halo can be
qualitatively understood using a spherical collapse model.  Following
Omukai (2001), we consider a spherical gas cloud embedded in a dark
halo.  We assume that the dynamics of the gas sphere is described by
\begin{equation}
\frac{d\rho_{\rm gas}}{dt} = \frac{\rho_{\rm gas}}{t_{\rm ff}},
\end{equation}
where $\rho_{\rm gas}$ is the gas density and the free fall time 
$t_{\rm ff}$ is given by
\begin{equation}
t_{\rm ff} = \sqrt{\frac{3\pi}{32 G \rho}}.
\end{equation}
We solve the energy equation 
\begin{equation}
\frac{de}{dt}= -p\frac{d}{dt}
\left(\frac{1}{\rho_{\rm gas}}\right)-\frac{\Lambda_{\rm cool}}{\rho_{\rm gas}},
\end{equation}
together with the chemical reactions and the cooling rate computed in a consistent manner 
as in our simulation. 
Specifically, the net cooling rate $\Lambda_{\rm cool}$ includes
cooling by molecular hydrogen $\Lambda_{\rm H_2}$, 
cooling by hydrogen and helium 

\begin{inlinefigure}
\resizebox{7cm}{!}{\includegraphics{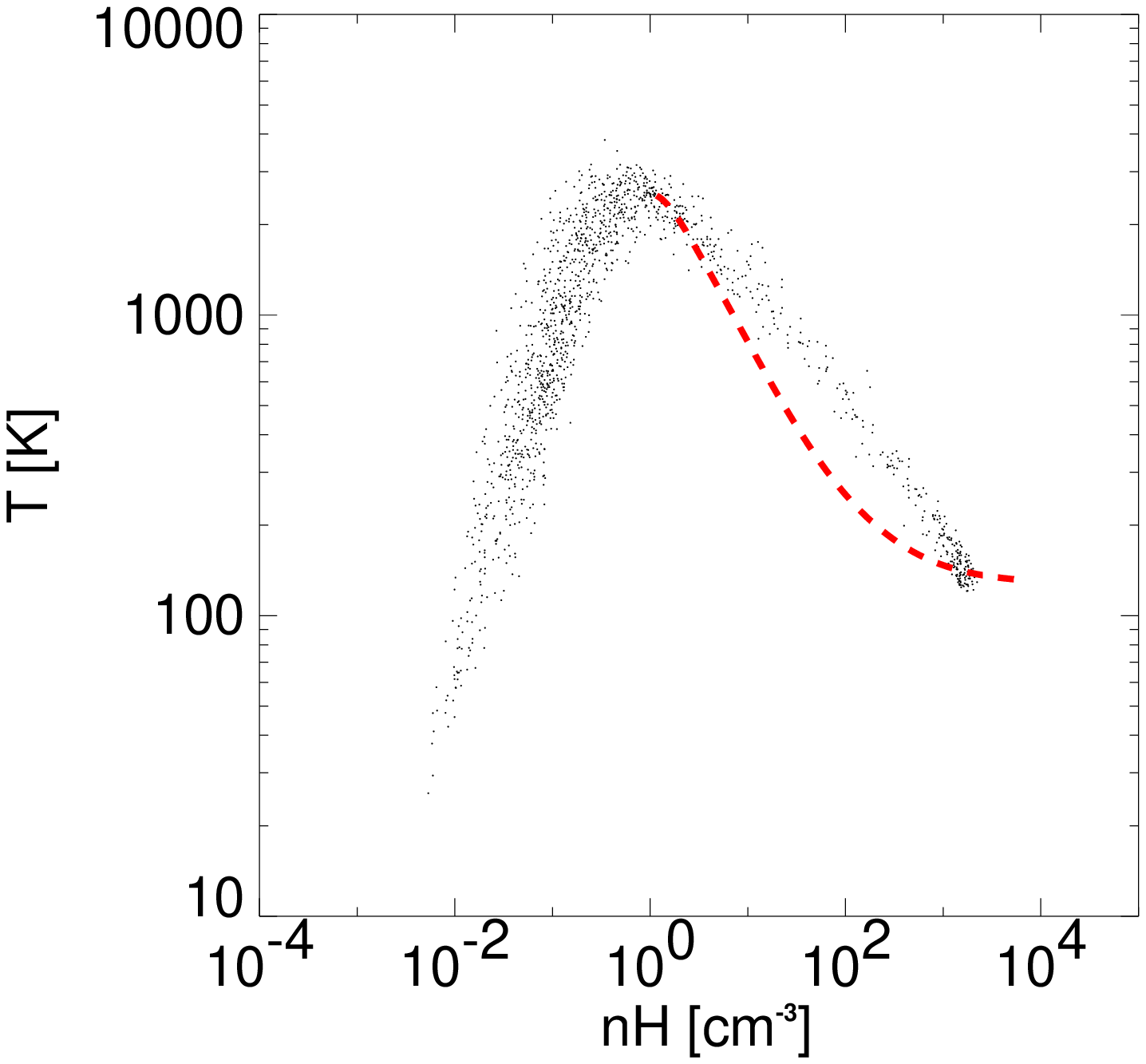}}
\caption{The distribution of gas particles within a 200 pc (physical) radius from the center
of the most massive halo in Run C1 at $z=20.7$ in the density - temperature plane. 
Densities are given in units of number density of
hydrogen atoms per cubic centimeters. The dashed line shows the
evolutionary track for spherically collapsing gas as described in section 6.
\label{plot7}}
\end{inlinefigure}
\\
atomic transitions $\Lambda_{\rm H, He}$,
and the inverse Compton cooling $\Lambda_{\rm Compton}$. 
Although the atomic line cooling is unimportant 
in the temperature range we consider, we include it for completeness.

For our purposes, we follow the evolution of the gas after it is
virialized. We take the initial temperature of the collapsing gas
cloud to be the virial temperature of the most massive halo in Run C1
at $z=23$, which has a mass of $6\times 10^5
h^{-1}M_{\odot}$ and a virial temperature $T_{\rm vir}=2300$ K.  
We follow the thermodynamic evolution of the gas from $z=23$
to $z=20.7$.
Figure \ref{plot7} shows the distribution of the halo gas particles in the
thermodynamic phase plane at $z=20.7$. We select the gas particles
within $200$ pc (physical) of the center of the halo.  The
trajectory computed by solving equations (7)-(9) is shown by the
dashed line.  It describes the evolution of the gas 
from $z=23$ to $z=20.7$ reasonably
well. Note that the cooling branch, appearing as dots in the right
portion of the plot, does not exactly represent the evolutionary track
of the gas particles.  The dots show the densities and temperatures of
the gas particles {\it at the output time}, $z=20.7$.
The cooled primordial gas piles up near the halo center with a
characteristic temperature $T \sim 100-200$K and number density
$n_{\rm H} \sim 10^3 - 10^4$cm$^{-3}$, in good agreement with the
prediction of the spherical collapse model.

\section{Effect of radiative feedback}

In the previous sections, we focused on the formation of the very
first objects in the absence of an external radiation field.  After the
first stars form, they emit photons in a broad energy range. Radiation
from the first stars affects not only the IGM in the vicinity of the
stars, but could also build up a background radiation field in certain
energy bands.  Photo-dissociation of molecular hydrogen due to
radiation in the Lyman-Werner (LW) bands (11.18eV - 13.6eV) is of
primary importance, because the LW radiation can easily penetrate into
the neutral IGM (Dekel \& Rees 1987; Haiman, Abel \& Rees 2000; Omukai
\& Nishi 1999; Glover \& Brand 2001).  We first model the influence of
LW radiation by including a uniform radiation background in the
optically thin limit, as in Machacek et al. (2001).  In section 7.3,
we also take into account gas self-shielding in the same set of
simulations and compare the results with those for the optically thin
cases.

\subsection{Uniform background radiation}

To begin, we adopt a constant radiation intensity in the LW band of
either $10^{-23}$ or $10^{-22}$ erg s$^{-1}$ cm$^{-2}$ Hz$^{-1}$
str$^{-1}$. 
We use the photo-dissociation reaction coefficient given in Abel et al. (1997),
\begin{equation}
k_{\rm diss}=1.38\times 10^9 J(h\nu =12.87 {\rm eV}).
\label{eq_diss}
\end{equation}
Hereafter, we describe the intensity by the conventional
normalization $J_{21}$. 
We adopt the values $J_{21}=0.01, 0.1$ by noting that the LW radiation with intensities $J_{21}\leq 0.001$ does not significantly affect the abundance of molecular
hydrogen in halos, whereas radiation with $J_{21} \geq 1.0$ will quickly 
dissociate hydrogen molecules. This can be easily seen by computing the dissociation 
time scale
\begin{equation}
t_{\rm diss}=k^{-1}_{\rm diss} \sim \frac{10^{12}}{J_{21}}\;\; {\rm sec}.
\end{equation}
During Run C1, we turned on the background
radiation at $z=24$, slightly before the formation of the first gas
cloud in the simulation box.  Although it might seem more consistent
to turn on the radiation only after the first object is formed, we
start it slightly earlier, in order to study the evolution of the gas
in the most massive halo as well.  Figure \ref{plot8} shows the
distribution of gas particles in the thermodynamic phase plane at
$z=20.7$.  The dashed lines in Figure \ref{plot8} are computed by
solving equations (7)-(9) including the LW radiation with $J_{\rm
21}=0.1$ and 0.01. 
For these cases we computed the evolution of the abundance of molecular hydrogen 
$f_{\rm H_2}$ for $J_{\rm 21}=0.1$ and 0.01.  
and then evaluated the gas cooling rate $\Lambda_{\rm H_2} (T, f_{\rm H_2})$.
For $J_{\rm 21}=0.01$, the cooling branch now
appears to lie on a shallower line than for the case with no radiation
(Figure 7). The analytic model again describes the feature reasonably
well. It is interesting that under the influence of LW radiation 
the characteristic
temperature of the gas clouds becomes higher than in the case with no
radiation.  Omukai (2001) argues that both the density and temperature
of gas clouds in the regime before they start to undergo run-away collapse
become higher for a stronger radiation field
and the characteristic Jeans mass
becomes {\it smaller}.  Although our simulation does not probe this
regime owing to lack of resolution, it is intriguing that the presence
of radiation affects the final mass of the collapsing gas clouds. Note
also that, in a cosmological context, even a slight delay of gas
cooling can make the subsequent evolution very different because of
the rapid formation of dark matter halos.

The case with $J_{21}$=0.1 is noticeably different from the other
two (Figure 7, 8) models.  Gas cooling and collapse are almost
entirely prevented when the LW radiation dissociates
molecular hydrogen.  For this case, we compute the H$_2$ formation and
dissociation timescales to be $t_{\rm H_2} = 30$ Myrs, $t_{\rm diss} =
3$ Myrs, respectively, with the latter being much smaller than the
former.  Thus, we expect the H$_2$ abundance to be close to the
equilibrium value.  The equilibrium H$_2$ abundance is then given by

\begin{inlinefigure}
\resizebox{6.5cm}{!}{\includegraphics{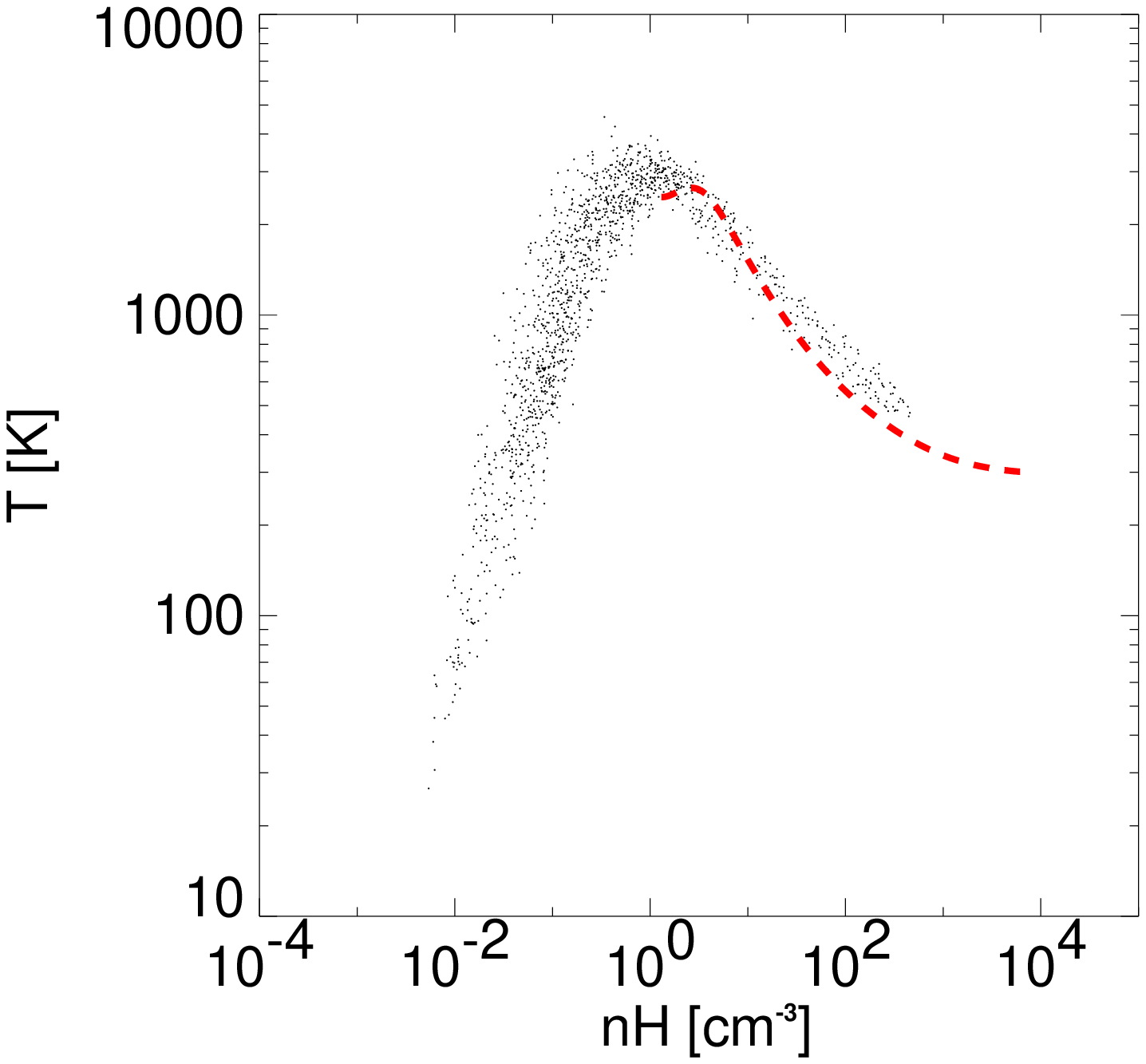}}\\
\resizebox{6.5cm}{!}{\includegraphics{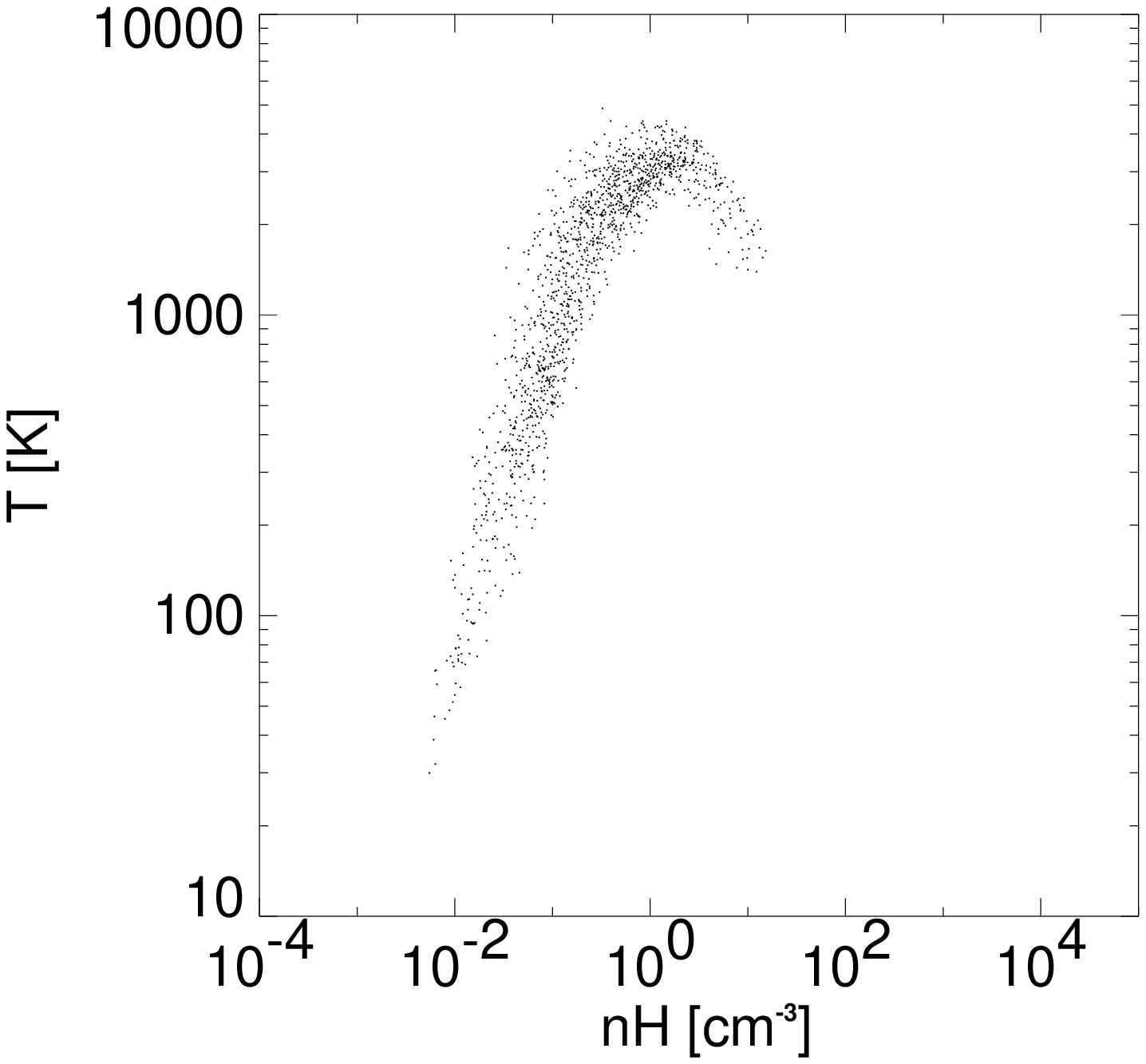}}
\caption{The distribution of gas in the density - temperature plane
 for the same halo as in Figure 7, but with Lyman-Werner background radiation
having flux $J_{21}=0.01$ (top) and  $J_{21}=0.1$ (bottom). 
Evolutionary track is not shown for the latter case, because the solution does not exist
for non-collapsing gas.
\label{plot8}}
\end{inlinefigure}
\\
\begin{equation}
n_{\rm H_2, eq} \approx \frac{k_{\rm H^{-}} n_{\rm H} n_{e}}{k_{\rm diss}},
\label{eq_eq}
\end{equation}
which yields the fraction $f_{\rm H_2} = 2\times 10^{-6}$ in the present case. It is nearly two orders of magnitude
smaller than the critical fraction needed to cool the gas (see Figure 2);
obviously, such gas cannot cool.

We generalize the above argument using the larger box simulation Run
A.  We turn on a uniform background radiation field with $J_{21} = 0.01$ at
$z=24$ and continue the run until $z=17$. We call this simulation
``Run A-r''.  Figure \ref{plot9} shows the mean molecular hydrogen
fraction against the virial temperature for the halos identified at
$z=17$ in this simulation.  As in Figure \ref{plot3}, the critical
molecular hydrogen fraction to cool the gas is shown by the solid
line, and the equilibrium H$_2$ abundance computed by equation (\ref{eq_eq}) is
indicated by the dashed line. For this plot, we computed the
equilibrium H$_2$ abundance by accounting for the fact that gas
densities in small halos are smaller than the universal baryon
fraction times the mean matter density within halos, because of
gas pressure (Loeb \& Barkana 2001).  The dashed line in Figure
\ref{plot9} thus appears to steepen toward the low temperature
end.  The agreement with the measured mean molecular fraction is quite
good, although the simulated halos show a large scatter at high
temperatures. 
Since the analytic model we adopted 

\begin{inlinefigure}
\resizebox{8cm}{!}{\includegraphics{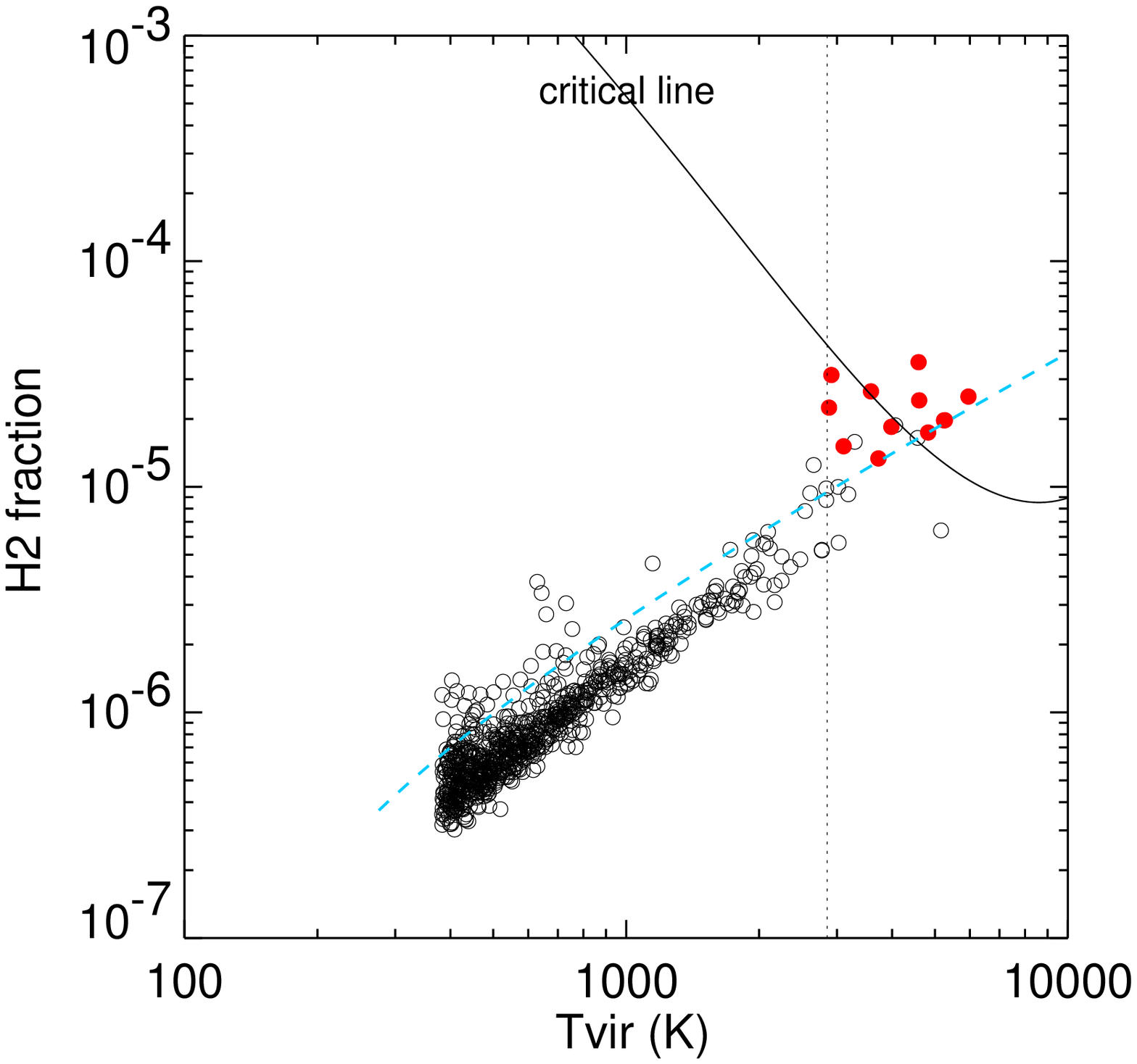}}
\caption{As for Figure 3, but for simulation Run A-r with a 
background radiation in the Lyman-Werner band with $J_{21}=0.01$.
\label{plot9}}
\end{inlinefigure}
\\
is expected to be accurate only to within some numerical
factor, we plot a factor of two smaller critical H$_{2}$ abundance in Figure \ref{plot9}
(solid line) than that in Figure 3. This brings the critical curve into
good agreement with the simulation results.  
In Figure \ref{plot9}, the critical temperature $T_{\rm cr}$ defined at the
point where the critical $f_{\rm H2}$ is equal to the equilibrium
H$_2$ abundance also agrees reasonably well with the actual minimum
temperature (vertical dotted line) found in our simulation.
 
Overall, we find that radiative effects are quite substantial. The
mean molecular hydrogen fractions drop by nearly an order of
magnitude for a radiation field with $J_{21}$ = 0.01. An order of
magnitude higher radiative flux will make the mean molecular fraction
even smaller and make primordial gas cooling very inefficient, as we
have seen in Figure \ref{plot8}.

\subsection{Self-shielding} 	

Although our simulation results in the previous section highlighted a
negative aspect of radiative feedback, the {\it true} importance of
this effect remains somewhat uncertain, because of our oversimplified
treatment of the background field.  The optically thin assumption
breaks down as dense gas clouds form, requiring that
self-shielding be taken into account. However, the strength of
self-shielding is a difficult question to address.  For a
static gas it is indeed significant, giving an effective shielding factor
$f_{\rm shield} \ll 1$ for molecular hydrogen column densities $N_{\rm
H_2} \gg 10^{14}$ cm$^{-2}$ (Drain \& Bertoldi 1996).  For a gas with
extremely large velocity gradients and disordered motion, the gas
remains nearly optically thin even for column densities $N_{\rm H_2}
\sim 10^{20-21}$cm$^{-2}$ (Glover \& Brand 2001). Since the full
treatment of three-dimensional radiative transfer for 76 LW lines,
even when only the lowest energy level transitions are included, is
practically intractable (see, however, Ricotti et al. 2001 for
one-dimensional calculations for a stationary gas), we adopt the
following approximate method to 
estimate the maximum effect of
self-shielding.  We consider only shielding by gas in
virialized regions and do not consider absorption by the IGM in
underdense

\begin{inlinefigure}
\resizebox{6.5cm}{!}{\includegraphics{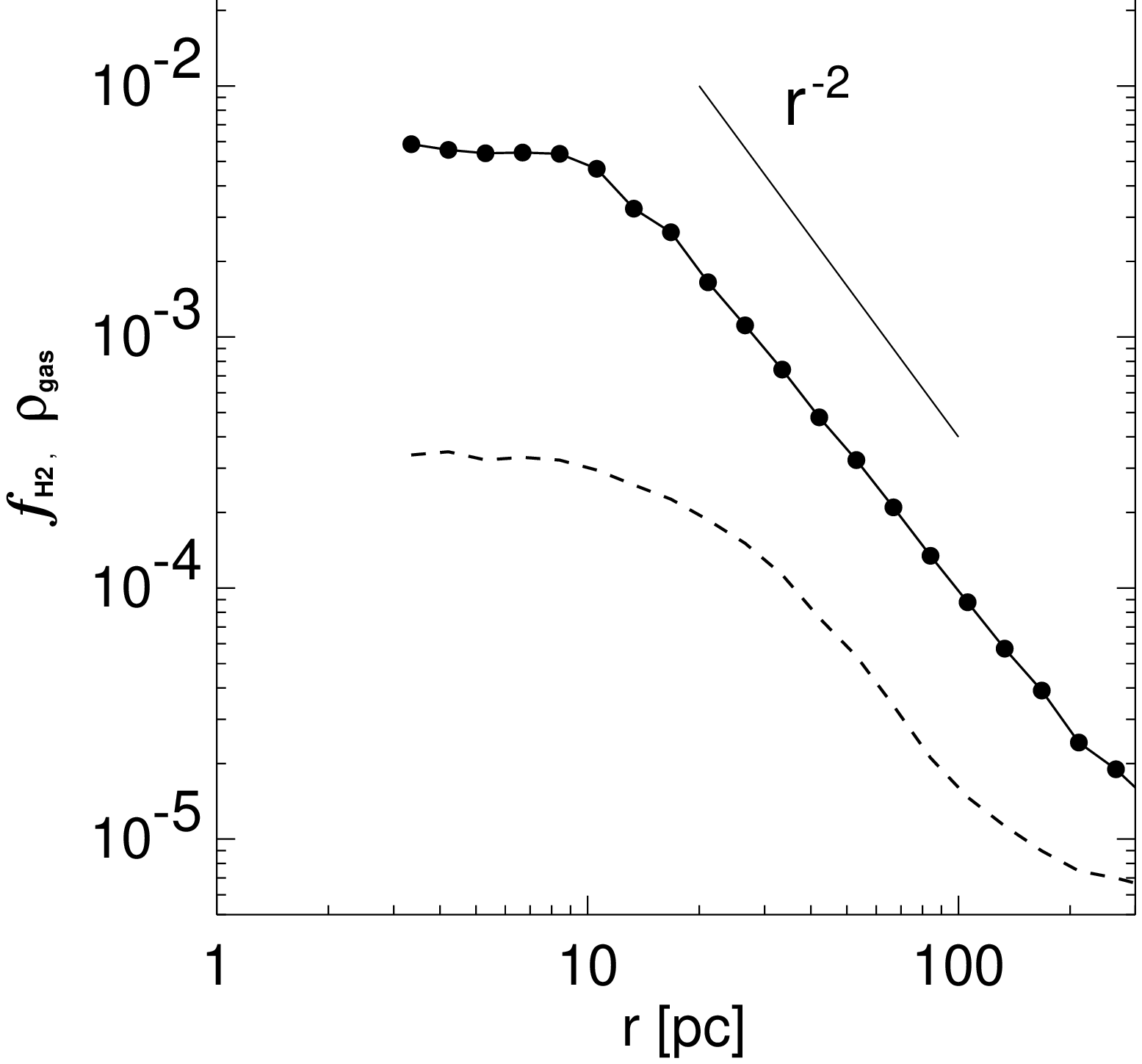}}\\
\vspace{0.5cm}\ 
\resizebox{6.5cm}{!}{\includegraphics{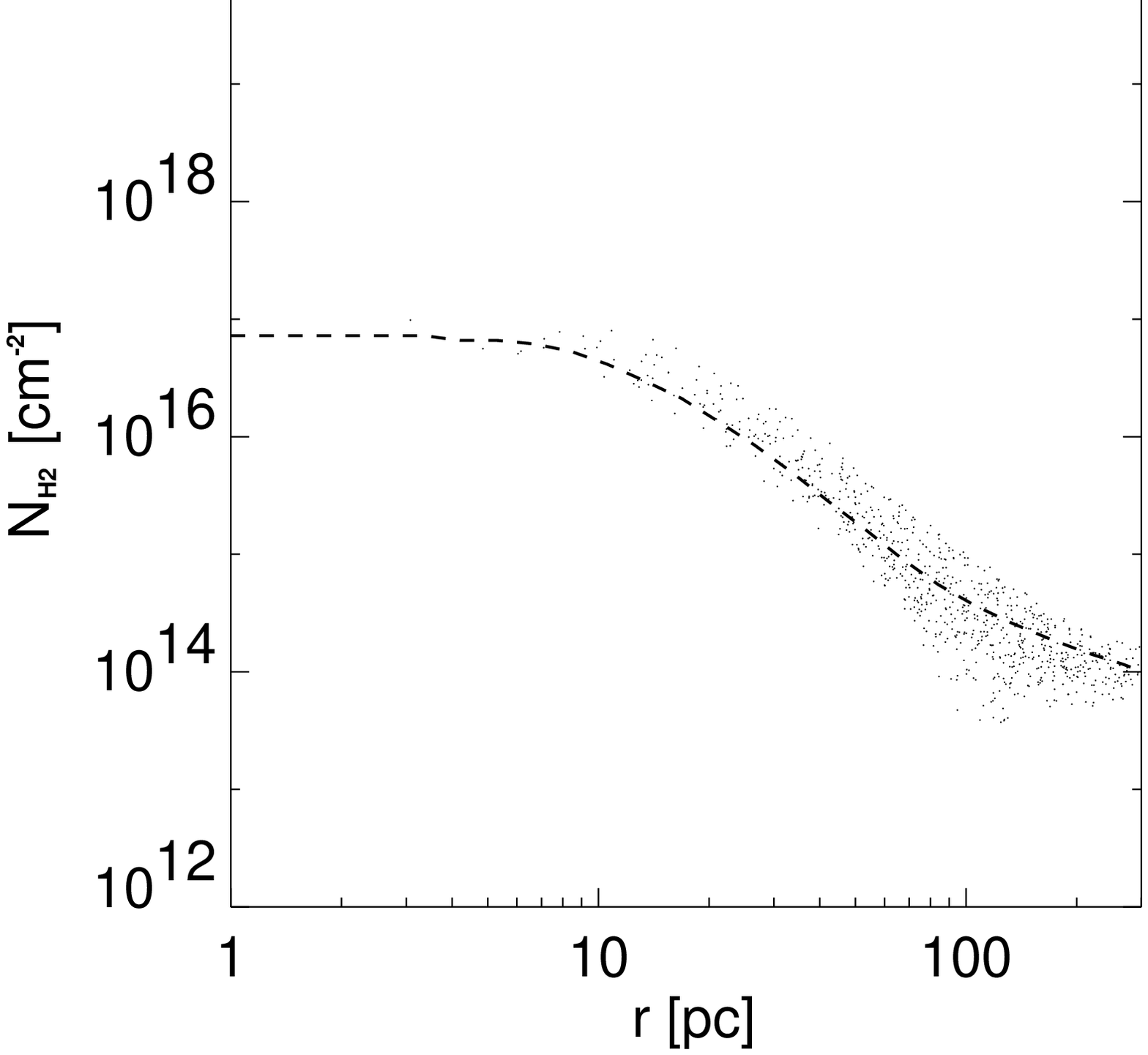}}
\caption{Top: The gas density profile (solid line with circles) 
and the molecular fraction profile (dashed line)
of the most massive halo in Run A at $z=25$.
An isothermal density profile is also shown by the solid line.
Bottom: The molecular hydrogen column density computed
from the profiles in the top panel (dashed line) 
and the column densities evaluated at the positions of the gas
particles are plotted against distance from the halo center.
\label{plot10}}
\end{inlinefigure}
\\
regions, assuming that the amount of residual
intergalactic H$_{2}$ is negligible. Although this is not true
initially, the intergalactic H$_{2}$ fraction quickly decreases after
the very first stars appear (Haiman, Abel \& Rees 2000).  On the other
hand, the so-called ``saw-tooth" modulation of background radiation
owing to neutral hydrogen Lyman series absorption (Haiman et al. 1997)
is substantial in the Lyman-Werner band because neutral hydrogen is
abundant at very high redshifts. Nevertheless, we do not consider the
evolution and attenuation of radiation and we fix the radiation
intensity as an input, rather than computing it consistently from
actual star formation. Thus the assigned intensity may be regarded as
that after the radiation is attenuated by intergalactic neutral
hydrogen.  We consider the evolution and modulation of the radiation
spectrum in section 8.

As for the simulations presented in section 7.1, we apply a
background radiation field in the LW band.  The radiation intensity at 
each position in the simulated region is computed by assuming that it
is attenuated through surrounding dense gas clouds.  We define a local
molecular hydrogen column density $N_{\rm H_{2}}$ in a consistent
manner employing the SPH formalism.  We use the local molecular
hydrogen abundance and 
\begin{inlinefigure}
\resizebox{8cm}{!}{\includegraphics{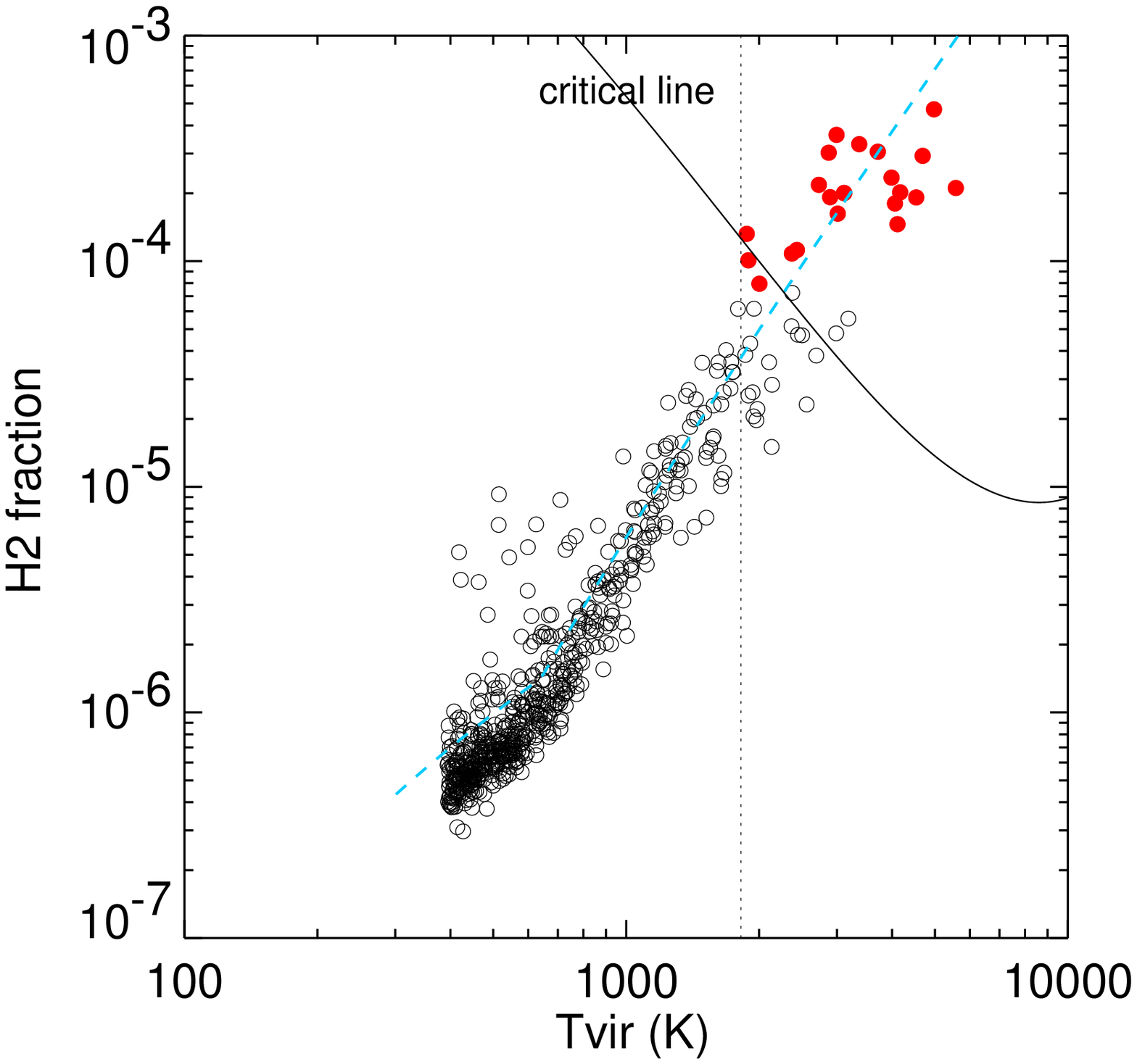}}
\caption{As for Figure 9, but for the simulation 
including the effect of gas self-shielding (Run A-s).
\label{plot11}}
\end{inlinefigure}
\\
density to obtain an estimate for the column density
around the $i$-th gas particle according to
\begin{equation}
N_{{\rm H_{2}}, i}=\int_{{\bf r}_{i}}^{r_{\max}} n_{\rm H_{2}} dl 
\end{equation}
where ${\bf r}_{i}$ is the position of the $i$-th gas particle and
$r_{\rm max}$ is the length scale we choose in evaluating the column
density.  In practice, we select an arbitrary line-of-sight and sum
the contributions from neighboring gas particles within $r_{\rm max}$
by projecting an SPH spline kernel for all the neighboring particles
whose volume intersects the sight-line.  We repeat this procedure
in six directions along x, y, and z axes centered at the position of the
$i$-th particle and take the {\it minimum} column density as a
conservative estimate.  We mention that our method is similar in
spirit to the local optical depth approximation of Gnedin \& Ostriker
(1997).  We have chosen the length scale $r_{\rm max}=100$ physical
parsec by noting that the virial radius of a halo with mass $10^6
h^{-1}M_{\odot}$ is just about 100 parsec.  There are not
significantly larger gas clumps than this scale in the simulated
region.  The local column density estimates are easily computed along
with other SPH variables with a small number of additional operations.

Figure \ref{plot10} shows the gas density and molecular hydrogen
fraction profiles for the most massive halo in Run A at $z=25$. The
gas density profile is very close to an isothermal density profile,
scaling as $\rho_{\rm gas} \propto r^{-2}$, except in the central 10
pc.  In the bottom panel, we compare the estimated column densities
at the position of each gas particle computed directly in the
simulation (dots) with the analytic estimate (dashed
line). 
We use the spherically averaged gas density and molecular
hydrogen fraction profiles shown in the top panel
and integrate from an arbitrary outer boundary as 
$\int_{\rm r}^{\rm r_{\rm out}} n_{\rm H} f_{\rm H_{2}} dr$,
to obtain the analytic estimate $N_{\rm H_{2}} (r)$.
The agreement is
quite good, assuring that our technique yields
accurate estimates for the column density.  Following Drain \& Bertoldi
(1996), 
we parameterize the shielding factor as
\begin{equation}
F_{\rm shield} = \min \left[ 1,  \left(\frac{N_{\rm H_{2}}}{10^{14} 
{\rm cm}^{-2}}\right)^{-3/4} \right ].
\label{eq_shield}
\end{equation}
The photo-dissociation reaction coefficient is then given by
\begin{equation}
k_{\rm diss}=1.38\times 10^9 J(h\nu =12.87 {\rm eV})F_{\rm shield}.
\label{eq_k31}
\end{equation}
The bottom panel of Figure \ref{plot10} shows that, at the halo
center, the column density is close to $10^{17}$ cm$^{-2}$, and hence
the radiation intensity is expected to be significantly reduced
according to equation (\ref{eq_shield}).  It should be
emphasized that the above expression for the shielding factor is
derived for a stationary gas, and thus the actual effect of
self-shielding could be substantially smaller because of gas velocity
gradients and disordered motions, as discussed in Machacek et
al. (2001).  The results using the above estimate can, therefore, be
regarded as describing the maximum possible effect of self-shielding.

We again use Run A. Similar to the optically thin case, we turn on a
background radiation field with $J_{21}=0.01$ at $z=24$, and compute
self-shielding factors for all the over-dense gas
particles (hence for virtually all the gas particles in virialized
regions).  We call this simulation ``Run A-s'' (for shielding).
Figure \ref{plot11} shows the mean molecular fraction against
virial temperature for halos in Run A-s at $z=17$.  The mean molecular
hydrogen fractions lie, with substantial scatter, on a steeper line
than for the optically thin case (compare with Figure
\ref{plot9}), indicating effective shielding.  For large halos,
indeed, the mean molecular hydrogen fractions are close to the values
we found in Run A (with no radiation, see Figure 3). On the other
hand, the gas in small halos is nearly optically thin, and their mean
molecular fractions are close to those found in the optically thin
limit (Figure \ref{plot9}).  We have found that we can quantify
this trend by computing an effective shielding factor
\begin{equation}
F_{\rm eff. shield} = F(N_{\rm H_2}=C f_{\rm H2} n_{\rm H} R_{\rm vir}),
\label{eq_effshield}
\end{equation}
where the function $F$ is defined in equation (\ref{eq_shield}),
$f_{\rm H2}$ is computed assuming no radiation, $n_{\rm H}$ is the
hydrogen number density taken to be 180 times the mean density, and
$R_{\rm vir}$ is the halo's virial radius. We have introduced a
constant factor $C$.  Choosing $C=0.2$ and computing the equilibrium
H$_2$ abundance from equation (\ref{eq_eq}) for the {\it
effectively attenuated} radiation flux using equation (\ref{eq_k31}),
we have obtained molecular fractions which agree very well with
the abundances found in the simulation.  The dashed line in Figure
\ref{plot11} shows the equilibrium H$_2$ abundance calculated in
this manner. A value somewhat smaller than unity was chosen for $C$ by
noting that a large fraction of the gas in the outer envelope of the
halos remains nearly optically thin, as can be inferred from Figure
\ref{plot10}.

\subsection{Minimum collapse mass under far UV radiation}

The net effect of photo-dissociating radiation is to raise
the minimum collapse mass scale for primordial gas cooling (Haiman
et al. 2000; Machacek et al. 2001; Wyse \& Abel 2003).
In Figure
\ref{plot12}, we plot the minimum mass of halos that host gas
clouds for 3 sets of simulations.
Although there are small
fluctuations, the minimum mass scales remain approximately constant,
at $5\times 10^5, 6.5\times 10^5,$ and $1\times 10^6 h^{-1}M_{\odot}$ for Run
A, Run A-s, and Run A-r, respectively.

\begin{inlinefigure}
\resizebox{8.5cm}{!}{\includegraphics{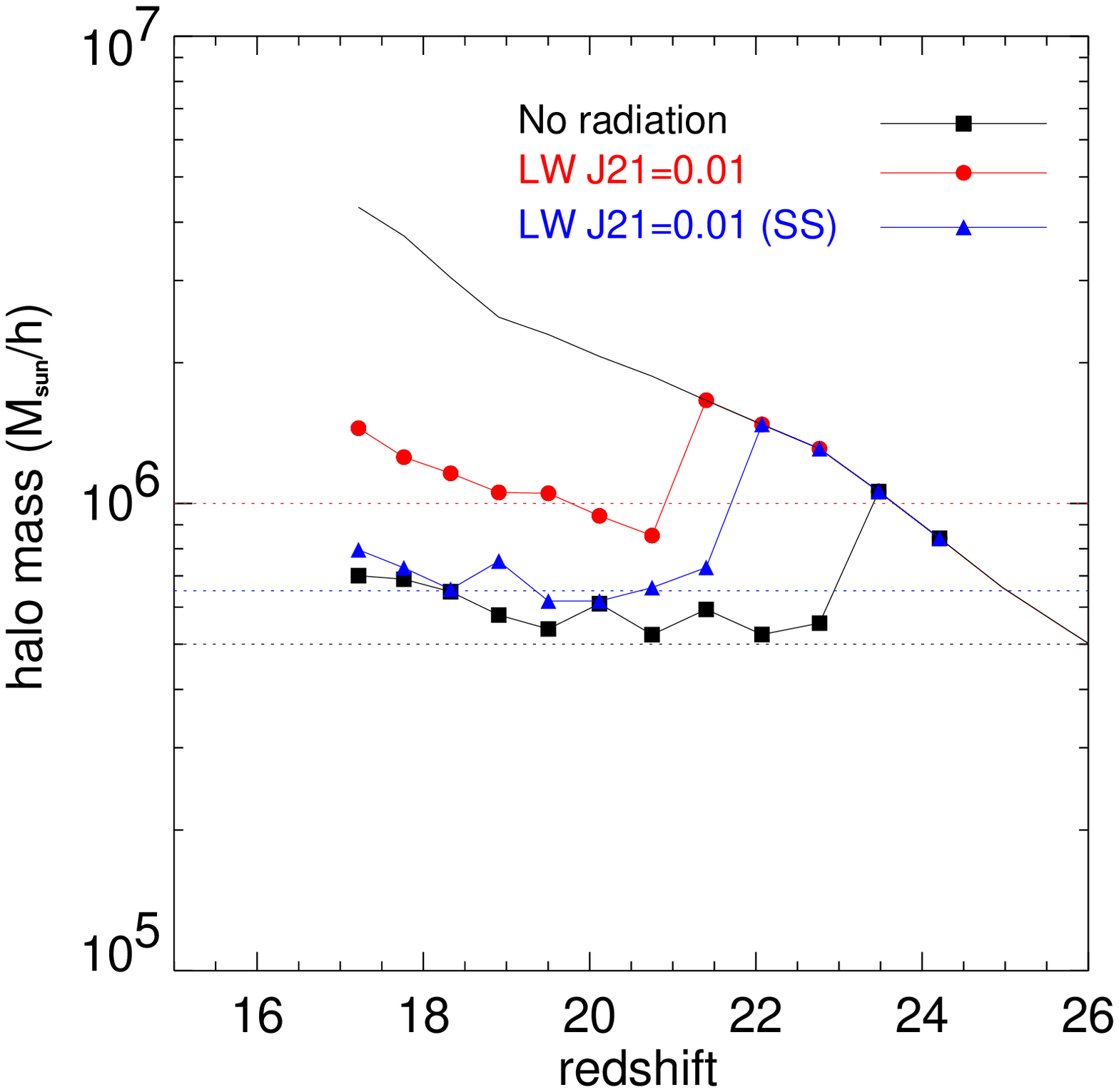}}
\caption{Minimum mass of the halos that host gas clouds in
Run A (no radiation case, squares), 
Run A-r (radiation flux $J_{21}=0.01$, circles), and
Run A-s (radiation flux $J_{21}=0.01$ with gas self-shielding
taken into account, triangles). The solid line shows the 
mass evolution of the most massive halo.
\label{plot12}}
\end{inlinefigure}

Figure \ref{plot13} summarizes our findings in this section.  
We compute the critical H$_{2}$ fraction and the asymtotic H$_{2}$ fraction
for no radiation case in the same manner as described in section 2.
We also show the equilibrium H$_{2}$ abundance for two cases with the LW radiation 
with $J_{21}=0.01$ (thick dotted line) and with $J_{21}=0.1$ (thin dotted line)
using equation (\ref{eq_eq}). Finally we compute the equilibrium abundance H$_{2}$
by taking the gas self-shielding into account using equations (14)-(16).
The effect of LW radiation is to reduce the fraction of
molecular hydrogen, $f_{\rm H_{2}}(T_{\rm vir})$, for a given virial temperature
$T_{\rm vir}$.  For optically thin radiation with $J_{21}=0.01$, $f_{\rm
H_{2}}(T_{\rm vir})$ is more than an order of magnitude smaller than in the case
with no radiation (compare the dashed line denoted as case (a) with the
thick dotted line denoted as (b)).  The case with maximal gas
self-shielding in our implementation lies in between these two cases
and bridges the low temperature ($\sim 600$ K) end of the optically
thin case and the high temperature ($\sim 4000$ K) end of the no
radiation case.  We expect that the true effect of gas
self-shielding 
should lie between these two extremes. The minimum collapse
mass scales for the three cases can be given, via the mass-temperature
relation in equation (1), by the crossing points of these three curves
with the critical molecular fraction shown by the solid line in Figure
\ref{plot13}.  As can be inferred from Figure \ref{plot13},
the minimum virial temperature for gas cloud formation by molecular
hydrogen cooling monotonically increases for increasing radiation
intensity. In the figure we also show the equilibrium
molecular hydrogen abundance for optically thin radiation with
$J_{21}=0.1$ by the thin dotted line. It does not cross the critical
$f_{\rm H_{2}}(T)$ curve, and thus, under such high intensity
radiation, molecular hydrogen cooling never becomes efficient in the
entire temperature range 
plotted. Note, however, that in large halos
with virial temperatures higher than $\simeq 7000$ K, the gas can cool
by atomic hydrogen transitions.  The overall cooling efficiency in
large halos is then dominated by atomic cooling and will

\begin{inlinefigure}
\resizebox{8.5cm}{!}{\includegraphics{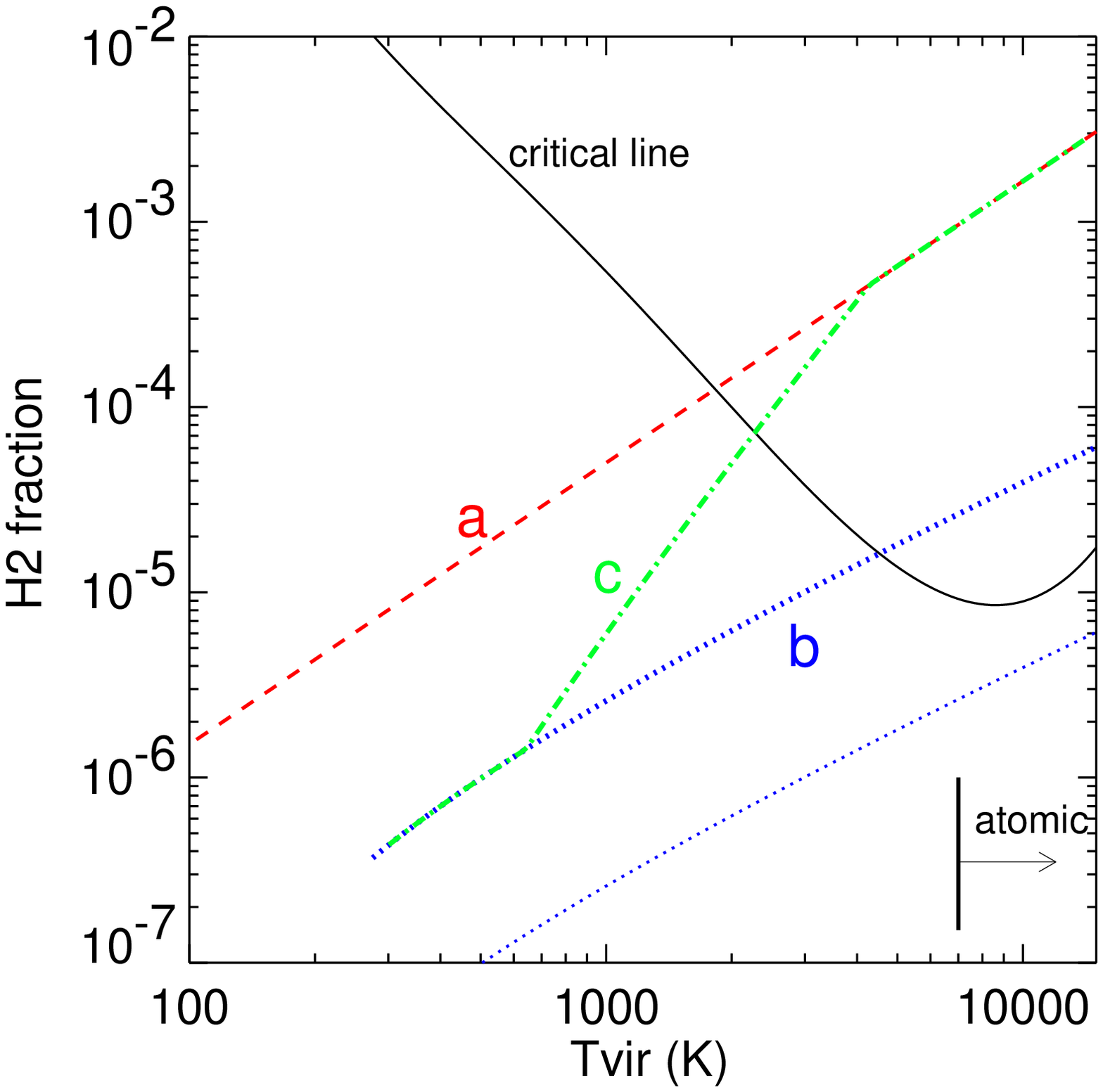}}
\caption{The critical molecular hydrogen fraction to cool the gas (solid line),
the H$_2$ fraction for (a) no radiation case (dashed line), 
(b) radiation with $J_{21}=0.01$ (thick dotted line), and
(c) radiation with $J_{21}=0.01$ with an effective self-shielding factor
taken into account (dot-dashed line). The thin dotted line is the equilibrium H$_2$ abundance
for a radiation flux $J_{21}=0.1$. These are computed at $z=17$. 
The vertical line shows the virial temperature above which atomic hydrogen line cooling becomes efficient.
\label{plot13}}
\end{inlinefigure}
\\
not be critically affected by the H$_2$ dissociating radiation regardless of
its intensity.
\subsection{Processes neglected}

Throughout this section we have considered radiation only in the
Lyman-Werner bands.  While photons with energy above 13.6eV are likely
to be completely absorbed by abundant neutral hydrogen in the IGM
surrounding the radiation source (but see below), those with energies
below the Lyman-Werner bands can easily penetrate into the IGM and so
the relevant processes involving these low energy photons should,
in principle, be taken into account.  The most important process in our
context is the photodetachment of H$^{-}$: ${\rm H}^{-} + \gamma
\rightarrow {\rm H} + e^{-}.$ Since H$^{-}$ catalyzes the formation of
molecular hydrogen, photodetachment by photons having sufficient
energy ($E > 0.755$ eV) could inhibit molecular hydrogen formation.
However, as discussed in Machacek et al. (2001), neglecting this
process does not affect our results because of the high densities in
the core regions where molecular hydrogen formation reactions (via
the H$^{-}$ channel) occur significantly faster than photodetachment
for the radiation intensities we used.

Haiman et al. (1996) and Kitayama et al. (2001) argue that a moderate
UV radiation field including photons with energy above 13.6eV can
promote molecular hydrogen formation and thus enhance primordial gas
cooling.  Since the overall strength of this {\it positive} effect
depends on the intensity and spectral shape of the radiation field in
a complicated manner, it is beyond the scope of the present paper to
analyze this in detail. It
also appears that such positive effects are appreciable only in a
restricted range of conditions.  We refer the reader to Haiman et
al. (1996) and Kitayama et al. (2001) for a discussion.

\section{Semi-analytic modeling of the early star formation}

Ideally, we wish to carry out large, high-resolution
$N$-body/hydrodynamic simulations with gas chemistry and radiation to
evolve structure formation and the cosmic radiation field together, in
a fully consistent manner.  Such simulations should employ large enough
volumes to contain a sufficient number of objects, while maintaining
at least the same resolution as the runs described here.  However,
computations such as this are still beyond our reach, in terms of
computational power.  We attempt to overcome this obstacle by applying
a semi-analytic method to a large dark matter $N$-body simulation
which can be carried out at a substantially lower cost.  We employ a
simulation with 324$^3$ CDM particles in a box of 1600 $h^{-1}$kpc
(Run DM in Table 1).  The simulation parameters were chosen such that
the mass per dark matter particle is $10^4 h^{-1}M_{\odot}$, allowing
us to resolve halos with masses $5\times 10^5 h^{-1}M_{\odot}$ by 50
particles.
In galaxy formation semi-analytic models that utilizes a halo merger tree
constructed from $N$-body simulations (e.g. Kauffmann et al. 1999), the smallest
halos are resolved by 10 simulation particles. Kauffmann et al. (1999)
quote that almost all the 10-particle halos identified in one output
are found as halos in later outputs in their high resolution simulations.   
Thus we justify that our choice of the simulation parameters allows  
us to robustly construct merger histories of halos with mass larger than 
$5\times 10^5 h^{-1}M_{\odot}$. 
We implement a set of simplified ``recipes'' to describe the
thermodynamic and chemical evolution of the gas in dark matter halos,
and calibrate the prescriptions against the numerical results
presented in the previous sections.  We emphasize that our model
significantly differs from, and improves upon, previous analytic
methods in which no dynamical information is incorporated.

\subsection{Building up the cosmic UV background}

Since the efficiency of gas cooling in halos is primarily determined
by the molecular hydrogen abundance, which is itself a function of the
background radiation field as well as the gas density and temperature,
we need to compute the evolution of the radiation flux coupled with
the formation and evolution of the first stars.  We compute the
frequency dependent radiation intensity at a given redshift from
\begin{equation}
J_{\nu}(z) = \int dz' c \frac{dt}{dz'} j_{\nu'}(z'),
\label{eq_Jevol}
\end{equation}
where $j_{\nu'}(z')$ is the total emissivity from all the sources at
redshift $z'$.  We need to account for the ``saw-tooth'' modulation of
the background radiation spectrum due to the Lyman-series absorption by
neutral hydrogen. (Note that hydrogen Lyman-$\alpha$ at 10.2 eV is
outside the range relevant to H$_2$ photo-dissociation.)  We follow
Haiman et al. (1997) and compute the saw-tooth modulation using a
screening approximation. Assuming that an effective screen due to
abundant neutral hydrogen blocks photons in the Lyman-series lines
from all sources at redshift above $z_{\rm max}$, we write equation
(\ref{eq_Jevol}) as

\begin{inlinefigure}
\resizebox{8.5cm}{!}{\includegraphics{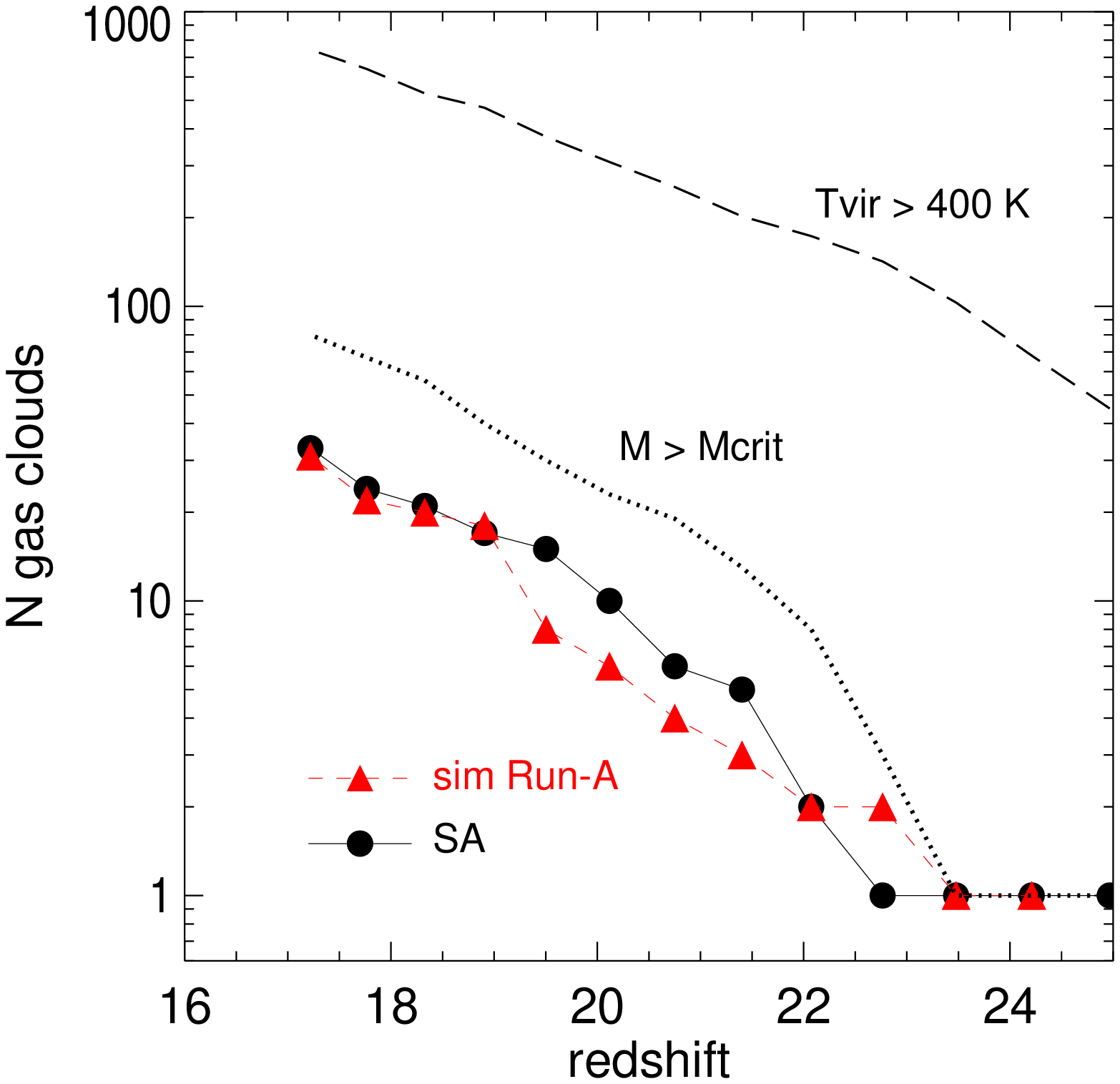}}
\caption{The number of star-forming gas clouds. The dashed line with triangles
shows the number of gas clouds found in our SPH simulation Run A. 
The solid line with filled circles shows the number of star-forming halos
computed by applying our semi-analytic model to the dark matter halos 
in Run A. The dotted line is the number of halos with mass greater than
the minimum collapse mass. It is about a few times larger than the number
of gas clouds in the simulation. Also, for reference, we plot the number of halos
with virial temperature greater than 400 K (long-dashed line);
it predicts a larger number of gas clouds by more than an order of magnitude.
\label{plot14}}
\end{inlinefigure}
\\
\begin{equation}
J_{\nu}(z) = \int_{z}^{z_{\rm max}} dz' c \frac{dt}{dz'} j_{\nu'}(z'),
\label{eq_Jevol2}
\end{equation}
where the maximum redshift for the $i$-th Lyman line at frequency
$\nu_{i}$ is given by
\begin{equation}
\frac{1+z_{\rm max}}{1+z_{\rm observe}}=\frac{\nu_{i}}{\nu_{\rm observe}}. 
\end{equation} 
The total source emissivity $j_{\nu'}(z')$ is computed by multiplying
the emissivity per single star by the number of active stars at
redshift $z'$.  We explore a specific model in which a single massive
Population III star is formed in each cold dense gas cloud.  We then
use the Pop III SED computed by Bromm, Kudritzki \& Loeb (2001),
assuming that the stars are more massive than 100 $M_{\odot}$.  For such
massive stars, the luminosity per unit stellar mass is $L(\nu) \approx
3\times 10^{21}$ erg s$^{-1}$ Hz$^{-1}$ $M_{\odot}^{-1}$ in the
Lyman-Werner band, with only a weak dependence on the stellar
mass.  Omukai \& Palla (2003) argue that Population III
stars with masses up to 600 $M_{\odot}$ could form by rapid
accretion, whereas Abel et al. (2002) claim, based on 
their simulation, that a reasonable estimate
for the maximum mass is 300 $M_{\odot}$.  In view of the uncertainty
in this quantity, we take the stellar mass to be a free 
parameter, keeping in mind that the total emissivity per
unit volume scales as the stellar mass, provided that one massive
star forms per gas cloud.  We also assume that the mean lifetime of
such massive stars is 3 million years (e.g. Schaerer 2002).
Then, the total emissivity is given by
\begin{equation}
j(z) = j_{*} N_{*}(z) = j_{*} N_{\rm clouds}(z),
\end{equation}
where $j_{*}$ is the emissivity per star and $N_{*}$ and $N_{\rm
clouds}(z)$ are the number of active stars and the number of
star-forming gas clouds at redshift $z$, respectively.

Our next task is to compute the number of star-forming gas clouds in
the simulation box.

\subsection{Gas cloud formation}

Using the outputs of Run DM, we locate dark halos in the same manner
as described in section 2 and construct a halo merger tree by tracing
the formation history of individual halos (see Yoshida et al. 2002 for
details).  We then employ a simplified, yet physically motivated,
prescription for the cooling of primordial gas within dark halos,
which is also based on the results presented in the previous sections.

For all the halos at a particular redshift, we judge whether the gas
within them can cool by determining if abundant numbers of 
hydrogen molecules have formed.  Specifically, we adopt the following
criteria for a halo to be star-forming: (1) the mean molecular hydrogen
fraction is greater than the critical molecular hydrogen fraction, and
(2) the recent mass growth rate of the halo is smaller than the critical
mass growth rate.  It might seem that an even simpler criterion $t_{\rm
dyn} < t_{\rm cool}$ suffices for halos at a given mass. While such
a model could work approximately if there is no evolving background
radiation, we prefer following the merger history explicitly. It is
important to specify {\it when} the gas in a halo cools, because the
coupling of the increasing radiation flux with star formation
makes the onset time of gas cooling critical.  The above cooling
criteria can be formulated as
\begin{equation}
f_{\rm H_2} (J, T_{\rm vir}, z) > f_{\rm H_2, crit},
\end{equation}
and
\begin{equation}
\left|\frac{dT}{dt}\right|_{\rm dyn.\; heat} < \left|\frac{dT}{dt}\right|_{\rm H_2\;\; cooling},
\end{equation}
where $f_{\rm H_2}(J, T_{\rm vir}, z)$ is a function of the background
radiation intensity $J$, the virial temperature $T_{\rm vir}$, and redshift $z$.  
We compute $f_{\rm H_2}(J, T_{\rm vir}, z)$ using the asymptotic molecular
fraction (equation [17] in Tegmark et al. [1997]) for no or negligible ($J_{21} < 0.001$) 
background radiation, while we use the equilibrium H$_{2}$ 
abundance using equation (12), (14), (15) and (16) for background radiation with $J_{21} \geq 0.001$.
The critical molecular fraction, $f_{\rm H_2, crit}$, is computed 
for the virial temperature $T_{\rm vir}$ and the density of the gas within a halo.
We approximate the instantaneous dynamical heating rate of a halo from 
its mass increase since the previous output time.  The choice of the
time interval for measuring the mass increase remains somewhat arbitrary.
Obviously, it should not exceed either the characteristic gas cooling
time or the dynamical time.  We take a conservative value of 3 Myrs for
the time interval, which satisfies these requirements for most of the
halos. 
It should be noted that our model does not trace the accumulation 
of cold dense gas. Cold gas clouds can gradually grow in mass 
over a long time scale. However, describing the evolution requires
specifying the gas density profile and local temperatures in a halo
at a given time. Instead of following such a complicated procedure, 
our model simply specifies the time when an enough amount of cold gas
($\sim M_{\rm Jeans}$) is accumulated. 
We applied this model to our Run A by discarding gas particles and
using only the dark matter particles with the particle mass scaled
appropriately. We followed the halo formation history and applied the
model to it. In Figure \ref{plot14} we plot the computed number of star-forming halos against
redshift. We compare it with the number of
the gas clouds found in Run A. The good agreement is encouraging. We
have also checked that not only do 
the total numbers of gas clouds agree,
but that there is nearly a one-to-one match between the star-forming halos
in our SPH simulation and those identified by our semi-analytic model. 
For reference, the number of halos with mass greater than $M_{\rm cr} =
5\times 10^5 h^{-1}M_{\odot}$ is also shown in Figure \ref{plot14}
(dotted line).  The simple minimum-mass model over-predicts the number
of star-forming clouds by up to a factor of three.  It is worth
mentioning that, in some previous works (e.g. Mackey et al. 2003), the
critical virial temperature for primordial gas cloud formation was
taken to be 400 K. In Figure \ref{plot14} we plot the number of
halos with virial temperature larger than 400 K.  It predicts more
than an order of magnitude larger (nearly two orders of magnitude at
$z > 22$) number of star-forming regions.  This over-estimate of the
number of star-forming regions would result in a substantial overestimate
for the star-formation rate and associated supernova rate.

\subsection{Pop III star formation at high redshift}

We now couple the formation of the first stars to the evolution of the
background FUV radiation so that we can predict the global star
formation rate.  Starting from the earliest output at $z=50$, we
identify star-forming regions in the manner described in section 8.2,
assuming initially that the background radiation intensity is zero. In
Run DM the first star-forming region appears at $z=32$.  We then
determine the total emissivity within the simulation box and compute a
frequency-dependent radiation flux at the {\it next} timestep
according to equation (\ref{eq_Jevol2}).  At every timestep, we
average the radiation intensities over frequency in the Lyman-Werner
band.  Then, a constant radiation field with the average intensity is
(assumed to be) applied. Namely, the mean intensity $\bar{J}(z)$ is
used to compute the molecular hydrogen fraction in halos using
equations (\ref{eq_eq}) and (\ref{eq_k31}), and then the cooling
criteria described in section 8.2 are checked for all the halos.
Figure \ref{plot15} shows the evolution of the background radiation
intensity $\bar{J}(z)$ and the spectrum of the processed radiation
field $J_{\nu}$ in the Lyman-Werner band at $z=21$. The evolution of
the radiation intensity is computed for two cases, one in the
optically thin limit (filled circles) and the other with gas
self-shielding (open squares), as described in section 7.  The
difference between the two cases becomes noticeable when the mean
radiation intensity exceeds $\sim 10^{-24}$ erg s$^{-1}$ cm$^{-2}$
Hz$^{-1}$ str$^{-1}$.  The dissociation time scale for radiation with
intensity below this level is $t_{\rm diss} = k^{-1}_{\rm diss} > 30$
Myrs, and thus its effect on gas cooling remains quite subtle even in
the optically thin case.  When the radiation intensity is above this
level, it dissociates hydrogen molecules quickly and suppresses
primordial gas cooling. It indeed acts more efficiently in the
optically thin case and quenches star-formation more strongly.
Suppressed star-formation makes the evolution of the background
radiation slower than in the model with self-shielding, as seen
in Figure \ref{plot15}.  Interestingly, our model predicts that the
mean background radiation intensity reaches a value 

\begin{inlinefigure}
\resizebox{8.5cm}{!}{\includegraphics{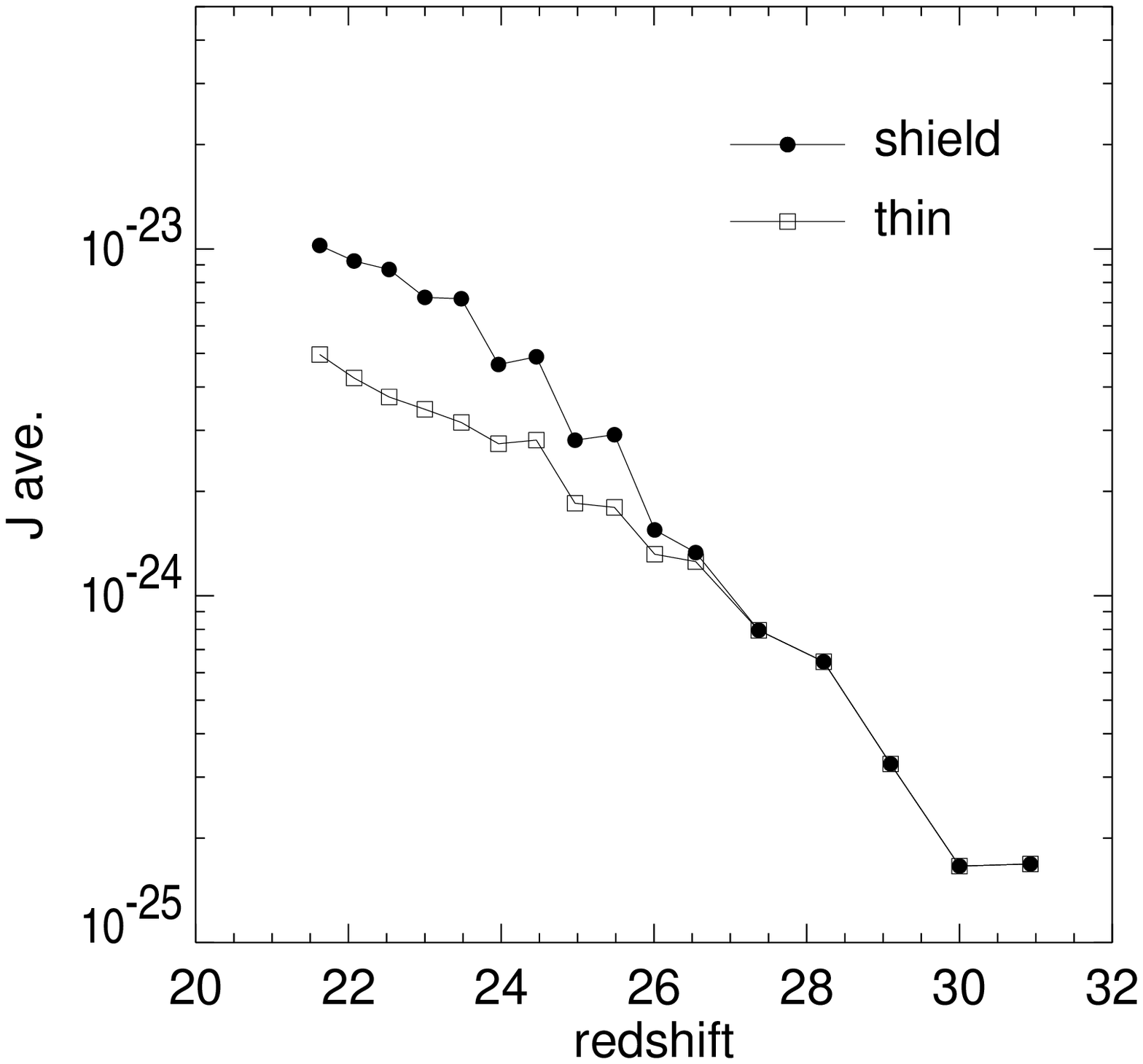}}\\
\resizebox{8.5cm}{!}{\includegraphics{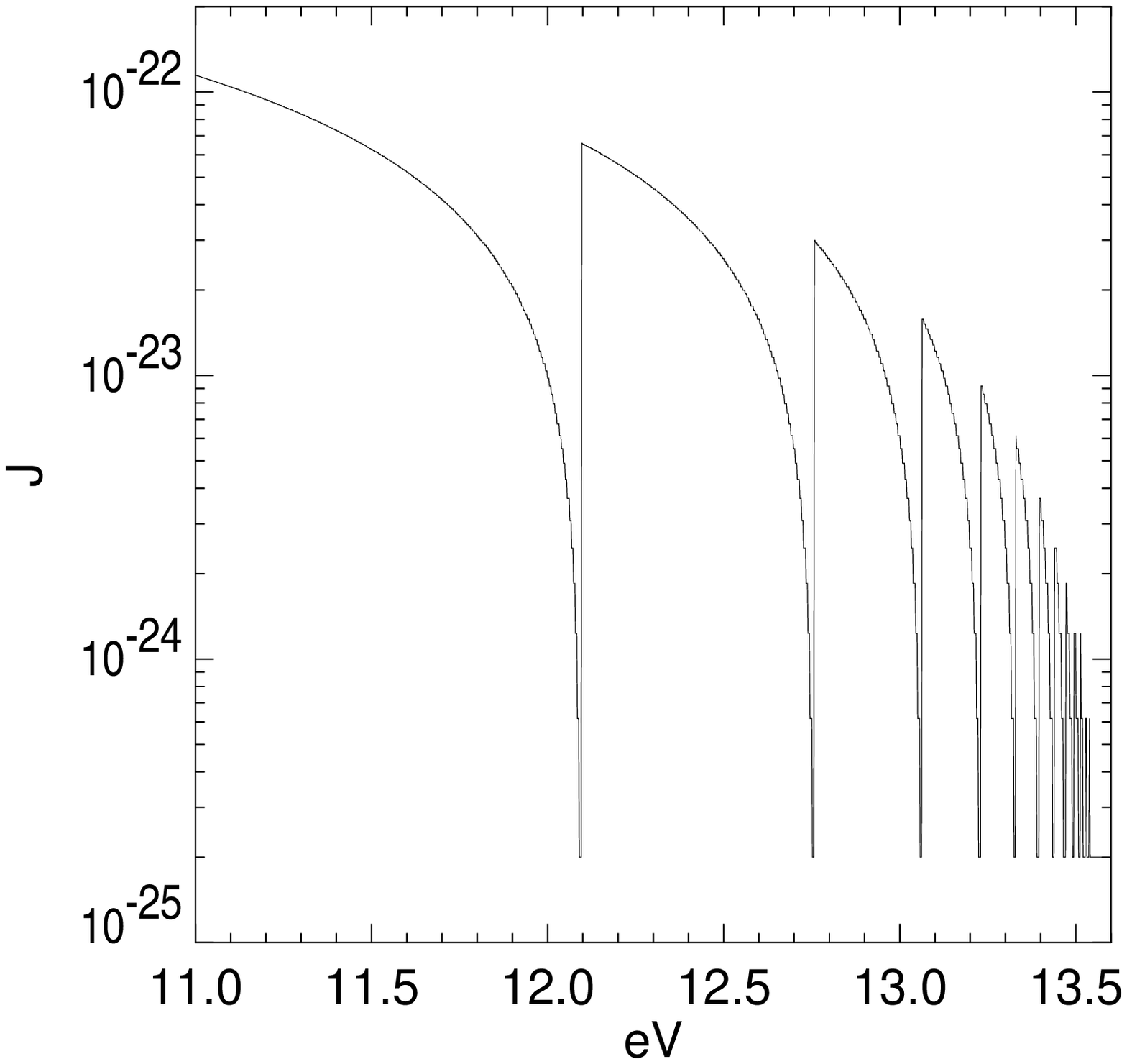}}
\caption{Top panel: The evolution of the background radiation intensity
for the two cases considered. 
The frequency averaged intensity in the range 11.18 - 13.6 eV is
shown.
Bottom panel: the spectrum at $z=21$ for the case with gas self-shielding. 
The optically thin case is quite similar except for the
overall intensity level.
\label{plot15}}
\end{inlinefigure}
\\
$10^{-23}$ erg s$^{-1}$
cm$^{-2}$ Hz$^{-1}$ str$^{-1}$ only at $z\sim 20$. This is because, as
is clearly seen in the bottom panel of Figure \ref{plot15}, the
hydrogen Lyman-series absorption causes a substantial intensity
decrease in the Lyman-Werner band. Without the absorption, the mean
intensity would have been more than an order of
magnitude higher.

As we advance to lower redshifts using our model, some halos grow to
have virial temperatures exceeding $T_{\rm crit. atomic}$ (see Figure
\ref{plot13}).  In principle, the gas should then cool via atomic hydrogen
transitions even if hydrogen molecules are completely
photo-dissociated.  It is conceivable that a large halo will be formed
through successive mergers in which molecular hydrogen cooling has
never become efficient (Hutchings et al. 2001).  However, we find that
all the massive halos ($> M_{\rm crit. atomic}$) have a progenitor in
which a star has already formed.  Our model

\begin{inlinefigure}
\resizebox{8.5cm}{!}{\includegraphics{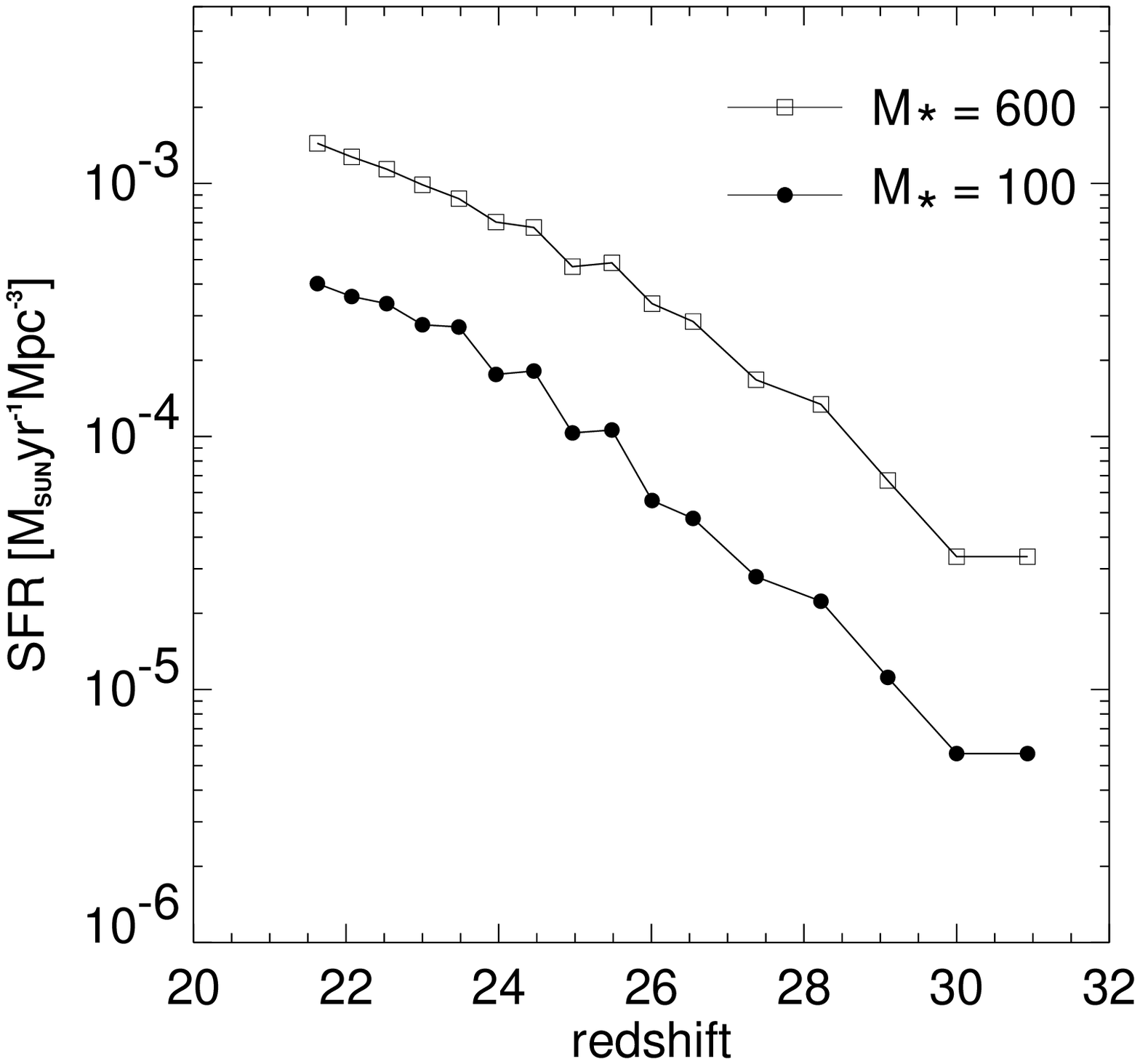}}
\caption{The comoving star formation rate density. 
Model M100 (filled circles) and model M600 (open squares).
\label{plot16}}
\end{inlinefigure}
\\
assumes that star
formation takes place only once in a halo and in its descendants
during the redshift range we consider.  This crudely mimics the strong
radiative feedback effect due to photo-dissociation and photo-heating
(Omukai \& Nishi 1999; Shapiro et al. 1997) in the vicinity of the
first stars. Since the physical time between $z=35$ and $=20$ is about
100 million years, the strong 
radiative feedback likely suppresses
subsequent star formation over a significant fraction of this period
(Yoshida, Abel \& Hernquist 2003), unless additional processes such as
metal dispersal by supernovae are invoked.
  
Finally, we use the number of stars formed to measure the conventional
comoving star formation rate (SFR) density,
$M_{\odot}$yr$^{-1}$Mpc$^{-3}$. Figure \ref{plot16} shows our model
prediction for the comoving Pop III star formation rate. We 
compare two cases by setting the mass of a Pop III star to be either
$M_{*}=100 M_{\odot}$ (filled circles, model M100) or $M_{*}=600
M_{\odot}$ (open squares, model M600).  We take a value of $600
M_{\odot}$ for the maximum Pop III star mass from Omukai \& Palla
(2003). For both cases we included gas self-shielding in the
model. (Thus model M100 in Figure \ref{plot16} is the result from the
same model as in Figure \ref{plot15}.)  The star-formation rate density
gradually increases from $z=30$ and reaches a value $\sim 10^{-3}
M_{\odot}$yr$^{-1}$Mpc$^{-3}$ at $z=20$. Note that in Figure
\ref{plot16} the predicted SFR can be approximately scaled by the
assumed stellar mass $M_{*}$ because of the straightforward unit
conversion we used.  In model M600, the star-formation rate is then
about 6 times larger than in model M100.  The coupling of
star-formation to the evolution of the background radiation causes
each to be regulated by the other (Wyse \& Abel 2003). As more stars
form, the radiation intensity rises, which suppresses primordial gas
cooling and hence the global star-formation rate. Then, the evolution
of the radiation field slows, maintaining the star formation rate at a
moderate level.  The flattening in the SFR for model M600 at $z < 24$
is due to this regulation. It is more prominent in model M600 than in
model M100 because the total emissivity per unit volume is larger in
model M600.

\begin{inlinefigure}
\resizebox{8.5cm}{!}{\includegraphics{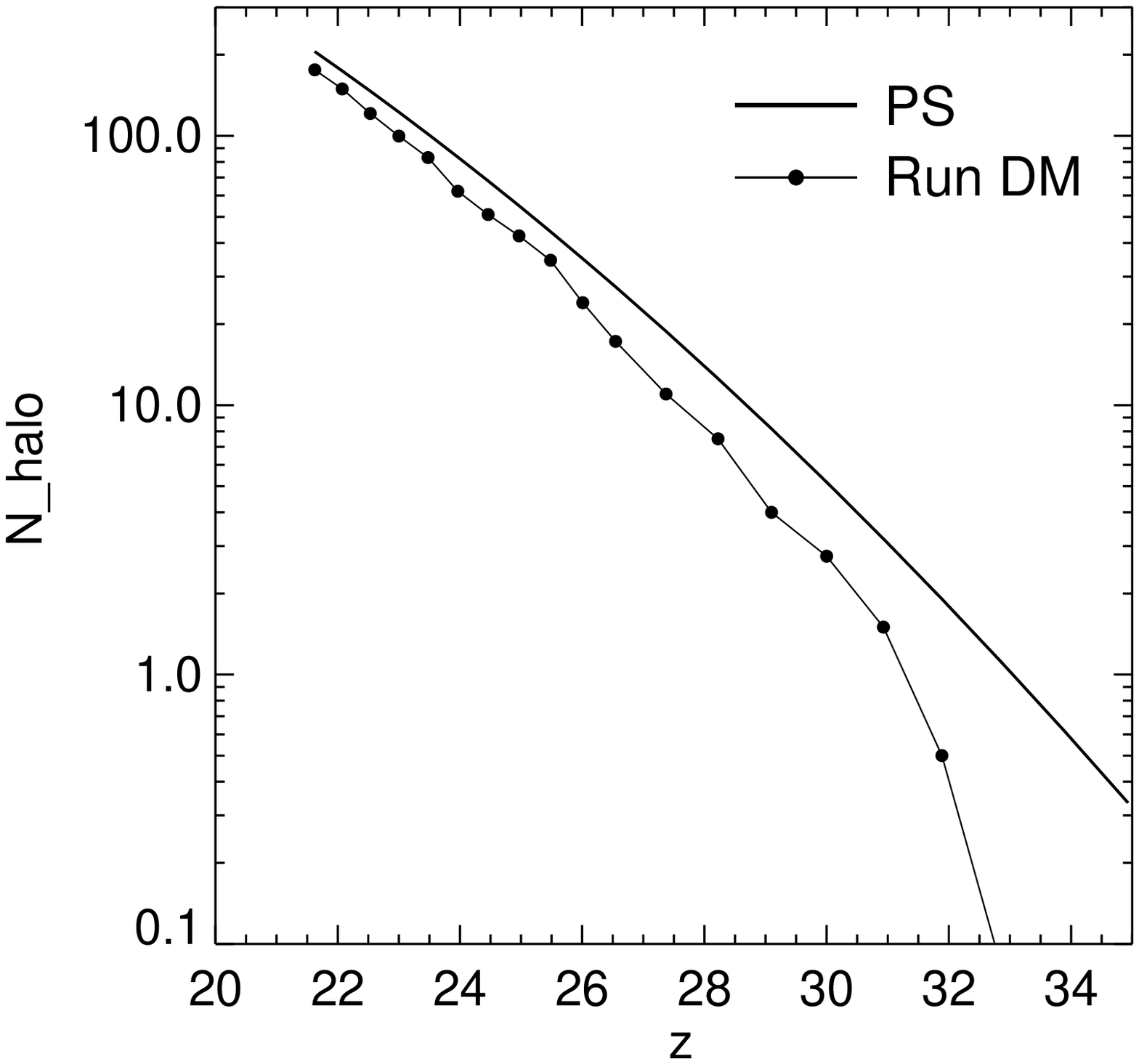}}
\caption{The number of halos with mass greater than $5\times 10^5 h^{-1}M_{\odot}$
per cubic mega-parsec (comoving) volume. 
\label{plot17}}
\end{inlinefigure}

We note that a characteristic feature of the Population III
star-formation calculated by Mackey et al. (2003) is not seen in our
result.  Their model assumes a sudden quenching of star-formation,
which produces a prominent ``cliff'' in their Figure 3.  The actual
regulation of the star-formation due to Lyman-Werner radiation occurs
in a more complex way as we have just described, and thus the
suppression of the Pop III star-formation is seen as a flattening of
the star-formation rate.
 
Intriguingly, the predicted SFR appears quite similar to the result of
the highest resolution simulation for PopIII stars by Ricotti et
al. (2002a), even though their simulations and model assumptions
differ from ours.  Ricotti et al. (2002b) presented extensive tests
using a series of simulations and examined the effect of various
parameters on the global star formation rate.  They showed that an
important parameter which influences the SFR at $z> 20$ is, indeed,
the size of the simulation volume.  Therefore, it is worth asking
whether the finite box size of our simulation affects the overall
results.  While the mass resolution of Run DM is enough for our
purposes (see the discussion at the beginning of this section), it is
not clear if the simulated volume (a cube of 1600$h^{-1}$ kpc on a
side) contains a fair sample of star-forming regions.  We address this
issue using the Press-Schechter halo mass function.  We have verified
that the mass function of the dark matter halos is reasonably well-fitted
by the Press-Schechter mass function in a mass range between $10^5 < 
M< 10^7 h^{-1}M_{\odot}$, and in a redshift range $20 < z <30$, in
agreement with Jang-Condell \& Hernquist (2001) at slightly lower
redshifts.

Since our analytical model relies on a one-star-per-halo assumption,
we use the total number of halos with mass greater than $M_{\rm crit}
= 5\times 10^5 h^{-1}M_{\odot}$ to test whether the abundance of large
halos is consistent with the Press-Schechter prediction.  Figure
\ref{plot17} shows the comoving number density of halos with mass
greater than $M_{\rm crit}$ computed from the Press-Schechter mass
function, in comparison with that found in Run DM.  They agree
reasonably 
well and the incomplete sampling of the halo mass function
due to the finite box size is appreciable only at $z > 30$. 
\begin{inlinefigure}
\resizebox{8.5cm}{!}{\includegraphics{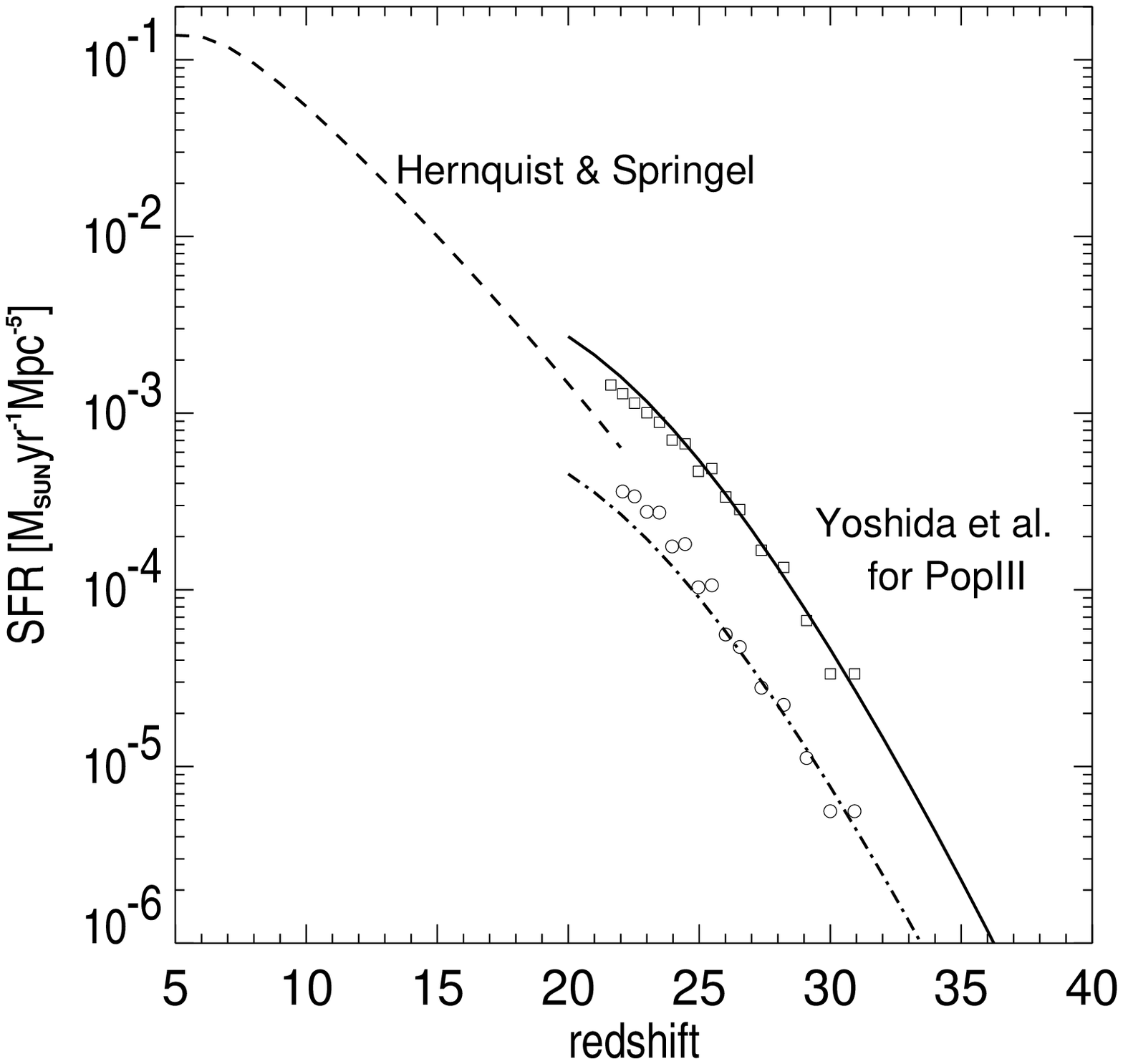}}
\caption{The cosmic star formation history.
We plot the star-formation rate density computed by 
our full semi-analytic model for $M_{*}$ = 100 $M_{\odot}$ (open circles)
and for $M_{*}$ = 600 $M_{\odot}$ (open squares).
The solid and dot-dashed lines are our simple functional fit (equation (23))
for the two cases. 
The analytic model prediction of the SFR by Hernquist \& Springel
(2003, their equation (2)) for the same $\Lambda$CDM cosmology is shown by the thick
dashed line. 
\label{plot18}}
\end{inlinefigure}
\\
Note also
that the small discrepancy remaining at lower redshifts ($20 < z < 30$)
may be partly due to inaccuracies in the analytic mass function itself.
Simulating a larger volume could make the discrepancy smaller at
$z>30$, but it would not significantly alter our results because
the star formation rate is not dominated by rare massive halos in 
our model. Therefore, we conclude that the level of agreement shown in
Figure \ref{plot17} over the relevant redshift range is satisfactory.

\subsection{The cosmic star formation history}

Springel \& Hernquist (2003a) and 
Hernquist \& Springel (2003) recently studied the star-formation
history of a $\Lambda$CDM universe using a set of numerical
simulations and an analytic model. They found that the evolution of
the star-formation rate can be well described by a simple functional
form.  Their simulations include only the ``normal-mode" of
star-formation in larger mass systems, where gas cooling occurs
via atomic hydrogen and helium transitions, and the regulation of
star-formation is governed by supernovae rather than radiation
(see Springel \& Hernquist 2003b).
Despite these differences, it is interesting to see how the
two modes of star-formation compare at high redshift.  In Figure
\ref{plot18}, we compare the star-formation at very high redshifts
($20 < z < 30$) computed from our model and that of Hernquist \&
Springel (2003) for $z < 20$.  
We show the SFR for two cases with
$M_{*}=100 M_{\odot}$ (open circles) and
$600 M_{\odot}$ (open squares). We also provide a simple functional fit
in a similar spirit to Springel \& Hernquist (2003a),
\begin{equation}
\rho_{*}(z) = \frac{1.4\times 10^{-5}M_{*}}
{1+(z-17)^{0.2} \exp(0.1(z-17)^{1.5})},
\end{equation}
in units of $M_{\odot}{\rm yr}^{-1}{\rm Mpc}^{-3}$.
The solid line in Figure \ref{plot18} shows equation (23) for $M_{*}=600 M_{\odot}$
and the dot-dashed line is for $M_{*}=100 M_{\odot}$.
This simple fit describes the results of our full semi-analytic model reasonably well,
as clearly seen in Figure \ref{plot18}.
The SFR of Hernquist \&
Springel (2003) begins just at $10^{-3} M_{\odot} {\rm yr}^{-1} {\rm
Mpc}^{_3}$, where our model prediction for the PopIII star formation
rate ends. The overall shapes of the SFR of the two modes appear remarkably similar,
although this may be just a coincidence since the physical mechanisms
governing star formation and its regulation are very different in
these two regimes.  We also note that feedback from the first
stars formed at $z \sim 20$ could affect
small (proto-)galaxy formation at $10 < z < 20$ by photo-heating, in
the same way as we argued in section 8.3.  Therefore it remains
unclear how the transition from Pop III to ``ordinary''
star-formation takes place. This issue clearly merits further
study.

\section{Discussions}

We have carried out cosmological simulations of primordial gas cloud
formation and determined the basic properties of the first baryonic
objects for the standard $\Lambda$CDM cosmology.  The minimum collapse
mass scale is set by the fraction of molecular hydrogen produced in
the primordial gas in dark matter halos, which is primarily determined
by the virial temperature of the system.  Our simulations reveal that
the merging process of dark halos disturbs the formation and evolution
of primordial gas clouds, and we detailed how such dynamical processes
delay the cooling and collapse of the gas.

We have also examined the impact of far ultra-violet radiation in the
Lyman-Werner bands on the formation of primordial gas clouds.  We have
followed the cooling and condensation of the gas within the most
massive halos in one of our simulations (Run C1) for a few cases with
a constant intensity radiation in the optically thin limit.  Due to
photo-dissociation of molecular hydrogen, gas cooling is suppressed
for radiation with intensity $J_{\rm 21} > 0.01$.  We showed that the
evolution of the gas cloud is qualitatively well-understood using a
spherical collapse model.  We then implemented a technique to compute
molecular hydrogen column densities in and around virialized regions
in the simulations and computed the gas self-shielding factor in its
maximum limit.  We found that, with the help of self-shielding,
primordial gas in large halos cools efficiently even when irradiated
by FUV radiation with intensity $J_{\rm 21} = 0.01$.  The overall
effect of the external FUV radiation is to raise the minimum halo mass
scale for efficient gas cooling. We quantified this effect using a
large volume simulation (Run A) and found that, in the optically thin
limit, the influence of FUV radiation can be formulated using an
equilibrium H$_{2}$ abundance.  Furthermore, we found that gas
self-shielding can also be described by defining an effective
shielding factor in a simple manner.  While our experiments in the two
extreme cases, the optically thin limit and with maximal gas
self-shielding, should bracket the true effect of gas self-shielding,
a more detailed study is needed to obtain an accurate estimate of this
process.  A promising approach may be to discard high velocity gas
elements in computing molecular hydrogen column densities (Simon
Glover, private communication).

Ideally, one would like to perform $N$-body hydrodynamic simulations
with gas chemistry and radiation using a large number of particles to
study the coupling of star formation to the evolution of the radiation
field.  Unfortunately, to simulate a volume equivalent to our dark
matter simulation Run DM with the same resolution as our Run A, would 
require at least $2\times 768^3$ particles.  As an
alternative, we developed a semi-analytic model for primordial gas
cooling and gas cloud formation.  By comparing our model predictions
with the results of SPH simulations, we found that the model
reproduces the number of primordial gas clouds in the simulated volume
very well. We then successfully applied it to outputs of a large
$N$-body simulation.  Assuming a simple star formation-law, we have
computed the star formation rate and the evolution of the background
radiation at high redshift ($20 < z < 35$).
We believe that our model can be used to reliably estimate the rate of
events associated with star formation, such as supernovae and gamma-ray
bursts.

While our simulations include the physics needed to study basic
properties of the first star-forming clouds, there are still a few
additional mechanisms that could be important.  An interesting
possibility proposed by Haiman et al. (2000) and Oh (2001) is that an
early X-ray background could increase the number of free electrons in
the IGM and promote the production of hydrogen molecules.  Analytical
estimates (Haiman et al. 2000) and numerical simulations (Machacek et
al. 2003) indicate, however, that the net effect is mild and the positive
feedback by an X-ray background does not entirely compensate the
negative feedback effect due to photo-dissociating Lyman-Werner
radiation unless unreasonably high X-ray intensities are assumed.  It
thus appears that our model prediction for the star-formation rate
should be robust, even without additional contributions from X-ray
sources.  It is probably more important to include direct radiative
transfer effects from the first stars.  In the present model, we only
crudely included the overall effect by quenching further star
formation locally after the first star is formed.  In our simulations
we find that many gas clouds are located close to one another, with
physical separations smaller than 10 kpc.  These star forming clouds
at high redshifts are, as we have shown, embedded in dark halos that
are strongly clustered (see Figure \ref{plot1}).  A single massive population
III star can ionize a large volume of the surrounding IGM and thus can
have considerable impact on the formation of the nearby gas clouds and
stars. In a forthcoming paper we will study direct radiative transfer
effects using ray-tracing simulations.

\acknowledgments

We thank Frank van den Bosch, Volker Springel, Edmund Bertschinger, 
Chung-Pei Ma, Volker Bromm, Tetsu Kitayama, and 
particularly Marie Machacek for fruitful discussions.  
We also thank the anonymous referee for giving us many constructive comments.
This work was supported in part by NSF grants
ACI 96-19019, AST 98-02568, AST 99-00877, and AST 00-71019.  The
simulations were performed at the Center for Parallel Astrophysical
Computing at the Harvard-Smithsonian Center for Astrophysics.

\section*{APPENDIX : Time Stepping in SPH with Non-equilibrium Chemistry}

We describe the time step criterion that incorporates the
non-equilibrium nature of the physics in our simulations.  
The code GADGET
employs an individual time step scheme and the usual time step
criterion for the $i$-th gas particle is given by the Courant condition
\begin{equation}
\Delta t_{i}=\frac{\alpha_{\rm cour} h_{i}}
	{h_{i}|(\nabla\cdot{\bf v})_{i}| + \max (c_{i},|{\bf v}_{i}|)(1+0.6 \alpha_{\rm visc})},
\end{equation}
where $h_{i}, c_{i}, {\bf v}_{i}$ have their usual meanings (see Springel et
al. 2001), $\alpha_{\rm visc}$ regulates the strength of the
artificial viscosity, and $\alpha_{\rm cour}$ is the Courant factor.
For each gas particle, we supplement this criterion with two additional
constraints so that the time step does not exceed the gas cooling time
and the characteristic chemical reaction time.
We monitor the cooling time 
\begin{equation}
\Delta t_{\rm cool} = e_{\rm tol} \frac{\;T\;}{\;\dot{T}\;}
\end{equation}
and use the rate of change of the electron number density to judge
the characteristic chemical reaction time
\begin{equation}
\Delta t_{\rm chem} = e_{\rm tol} \frac{\;n_{\rm elec}\;}{\;\dot{n}_{\rm elec}\;}.
\end{equation}
We set the tolerance parameter $e_{\rm tol}=0.1$ throughout the simulations.

\end{document}